\documentclass[3p,preprint,12pt]{elsarticle}

\usepackage{amsfonts,amsmath,amssymb}

\usepackage{natbib}

\usepackage[caption = false]{subfig}
\usepackage{hyperref}
\hypersetup{
    colorlinks=true,
    linkcolor=blue,
    filecolor=magenta,      
    urlcolor=cyan,
}
\usepackage{makecell,bbm}
\usepackage{multirow}
\newcommand\MyBox[2]{
  \fbox{\lower0.75cm
    \vbox to 1.7cm{\vfil
      \hbox to 1.7cm{\hfil\parbox{1.4cm}{#1\\#2}\hfil}
      \vfil}%
  }%
}

\usepackage{capt-of}

\usepackage{algorithmic}
\usepackage[Algorithm,ruled]{algorithm}

\usepackage{graphicx}
\usepackage{float}
\usepackage[utf8]{inputenc}
\usepackage{bm}
\usepackage{pdfpages}

\DeclareMathOperator*{\argmax}{arg\,max}

\newcommand{\dens}{\texttt{p}}
\def\thml{\hat{\theta}_{\rm ML}}

\newcommand{\prob}[1]{\mathbb{P}\left(#1\right)}
\def\by{y}
\def\bx{x}
\def\bz{z}
\def\bbeta{\beta}
\def\like{{\cal L}}
\def\llike{{\cal LL}}

\def\iid{\mathop{\sim}_{\rm i.i.d.}}

\def\ximis{\bx_{i,{\rm mis}}}
\def\xiobs{\bx_{i,{\rm obs}}}
\def\xmis{\bx_{{\rm mis}}}
\def\xobs{\bx_{{\rm obs}}}
\def\xa{\bx^{(\rm mis)}}
\def\xb{\bx^{({\rm obs})}}

\begin{document}
\begin{frontmatter}
  \title{Logistic Regression with Missing Covariates -- Parameter Estimation, Model Selection and Prediction within a Joint-Modeling Framework}
  \author[1]{Wei Jiang\corref{cor1}}
  \ead{wei.jiang@polytechnique.edu}
  \author[1]{Julie Josse}
    \author[1]{Marc Lavielle}
      \author[2]{TraumaBase Group}
      
       \cortext[cor1]{Corresponding author}
        \address[1]{Inria XPOP and CMAP, École Polytechnique, France}
      \address[2]{Hôpital Beaujon, APHP, France}

\begin{abstract}
Logistic regression is a common classification method in supervised learning. Surprisingly, there are very few solutions for performing logistic regression with missing values in the covariates. We suggest a complete approach  based on a stochastic approximation version of the EM algorithm to do statistical inference with missing values including the estimation of the parameters and their variance, derivation of confidence intervals and a model selection procedure. 
We also tackle the problem of prediction for new observations (on a test set) with missing covariate data.
The methodology is computationally efficient, and its good coverage and variable selection properties are demonstrated in a simulation study where we contrast its performances  to other methods. For instance, the popular approach of multiple imputation by chained equations can lead to estimates that exhibit meaningfully greater biases than the proposed approach. We then illustrate the method on a dataset of severely traumatized patients from Paris hospitals to predict the occurrence of hemorrhagic shock, a leading cause of early preventable death in severe trauma cases. The aim is to consolidate the current red flag procedure, a binary alert identifying patients with a high risk of severe hemorrhage. The methodology is implemented  in the R package misaem. 
\end{abstract}

\end{frontmatter}
\section{Introduction}\label{sectionintro}

Missing data exist in almost all areas of empirical research. There are various reasons why missing data may occur, including 
survey non-response, unavailability of measurements, and lost data.
One popular approach to handle missing values consists in modifying an estimation process so that it can be applied to incomplete data. For example, one can use the EM algorithm \citep{dempster1977} to obtain the maximum likelihood estimate (MLE) despite missing values, accompanied by a supplemented EM algorithm (SEM) \citep{sem}  or Louis' formula \citep{louis} for their variance.  
This strategy is valid under missing at random (MAR) mechanisms \citep{little_rubin,MAR}, in which the missingness of data is independent of the missing values, given the observed data.
Even though this approach is perfectly suited to specific inference problems with missing values, there are few solutions or implementations available, even for simple models such as logistic regression, the focus of this paper.

One explanation is that the expectation step of the EM algorithm often involves unfeasible computations. In the framework of generalized linear models \citet{ibrahim1999_MonteCarlo,compare_glm}, suggested to use a Monte Carlo EM (MCEM) algorithm \citep{mcemWeiTanner,EMMcLachlan}, replacing the integral by its empirical sum using Monte Carlo sampling. \citet{ibrahim1999_MonteCarlo} also estimated the variance using a Monte Carlo version of Louis' formula by Gibbs sampling with an adaptive rejection sampling scheme \citep{gilks1992adarej}.
However, their approach is computationally expensive and they considered an implementation only for monotone patterns of missing values, or for missing values only in two variables in a dataset. 


\textcolor{black}{In this paper, we develop 
 a stochastic approximation version of the EM algorithm (SAEM) \citep{lavielle:hal-01122873}, based on Metropolis-Hastings sampling, to perform statistical inference for logistic regression with incomplete data, where the missing data can be anywhere in the covariates.} 
SAEM uses a stochastic approximation
procedure to estimate the conditional expectation of the complete-data likelihood, instead of generating a large number of Monte Carlo samples which lead to an undeniable computational advantage over MCEM as illustrated in the simulation studies. 
In addition, it allows for model selection using criterion based on a penalized version of the observed-data likelihood.  This latter characteristic is very useful in practice, as few methods are available to select a model when there are missing values. For example, \citet{missaic,missaic2} suggested an approximation of AIC, while \citet{gic_af} defined generalized information criteria and  in the framework of imputation \citet{mirl} proposed to combine penalized regression techniques with multiple imputation and stability selection.
Besides aiming at maximizing the MLE for observed data, \citet{chow1979look, fung1989} studied the linear discriminant function for logistic regression, using pairs of observed values in columns to calculate the covariance matrix. Note that another solution is to use Laplace approximation to compute integrals, however, this  approximation linearizes the likelihood function by differentiation whereas SAEM performs exactly the inference.

This paper proceeds as follows: In Section \ref{sectionmot} we describe the motivation for this work, the TraumaBase\footnote{\url{http://www.traumabase.eu/}} project based on a French multicenter prospective Trauma Registry. Section \ref{sectionnote} presents the assumptions and notation used throughout this paper.
In Section \ref{sectionsaem}, we derive an algorithm SAEM to obtain the maximum likelihood estimate of parameters in an logistic regression model for continuous covariate data, under the MAR mechanism and a general pattern of missing data. Following the estimation of parameters, we present how to estimate the Fisher information matrix using a Monte Carlo version of Louis' formula. 
Section \ref{sectionmodselect} describes the model selection scheme based on a Bayesian information criterion (BIC) with missing values.
In addition, we propose an approach to perform prediction for a new iobservation with missing values.
Section \ref{sectionsimu} presents a simulation study where the proposed approach is compared to alternative methods such as multiple imputation  \citep{mi}, which may suffer from greater biases than the proposed approach and under-coverage. In Section \ref{sectiontrauma}, we apply the newly developed approach to predict the occurrence of hemorrhagic shock in patients with blunt trauma to the TraumaBase dataset, where it is crucial to efficiently manage missing data because the percentage of missing data varies from 0 to 60\% depending on the variables. Compared to the predictions made by emergency doctors, the results are improved with SAEM. Finally, Section \ref{sectiondisc} concludes this work and provides a discussion.

Our contribution is to provide users the ability to perform logistic regression with missing values within a joint-modeling methodological framework that combines computational efficiency and a sound theoretical foundation. 
The methodology presented in this article is implemented as an R \citep{softR} package \textit{misaem} \citep{misaem}, available in CRAN. The code to reproduce all the experiment is also provided in GitHub \citep{github}.

\section{Medical emergency}\label{sectionmot}

Our work is motivated by a collaboration 
with the TraumaBase group at APHP (Public Assistance - Hospitals of Paris), which is dedicated to the management of severely traumatized patients.

Major trauma refers to injuries that endanger a person's life or functional integrity. The WHO has recently shown that major trauma - road accidents, interpersonal violence, falls, etc. - are a worldwide public health challenge and a major source of mortality (first cause in the age group 16-45) and disability (2nd cause) in the world \citep {20171260}. The two leading causes of death are hemorrhagic shock and traumatic brain injury.


The path of a traumatized patient takes place in several stages: from the accident site where he is taken care of by the ambulance to the transfer to intensive care unit for immediate interventions and finally comprehensive care at the hospital. 
Using a pre hospital patient's records, we aim to establish models to predict the risk of severe hemorrhage to prepare an appropriate response upon arrival at the trauma center; e.g., massive transfusion protocol and/or immediate haemostatic procedures. 

Due to the highly stressful and multi-player environments involved, evidence suggests that patient management -- even in mature trauma
systems -- often exceeds acceptable time frames \citep{hamada2014evaluation}. In addition, discrepancies may be observed between the diagnoses made by emergency doctors in the ambulance,  and those made
when the patient arrives at the trauma center \citep{Hamada2015EuropeanTG}. These discrepancies can  result in poor outcomes such as inadequate hemorrhage control or delayed transfusion.

To improve decision-making and patient care,
15 French trauma centers have collaborated to collect detailed high-quality clinical
data from the accident scene, to the
hospital. 
The resulting database, TraumaBase, is a multicenter prospective trauma registry that is continually updated and now has data from more than 7,000 trauma cases.  The granularity of collected data (with more than 250 variables) makes this dataset unique in Europe.  However, the data from multiple sources, are highly heterogeneous, and are often missing, which makes modeling challenging.

In this paper, we focus on performing logistic regression with missing values to help propose an innovative response to the public health challenge of major trauma.

\section{Assumptions and notation}\label{sectionnote}


Let $(\by,\bx)$ be the observed data with $\by=(y_i , 1\leq i \leq n)$  an $n$-vector of binary responses coded with $\{0, 1\}$ and $\bx= (x_{ij}, 1\leq i \leq n, 1 \leq j \leq p)$ a  $n\times p$ matrix of covariates,  where $x_{ij}$ takes its values in $\mathbb{R}$. 
The logistic regression model for binary classification can be written as:   
\begin{equation} \label{regmodel}
\prob{y_i=1|\bx_i;\bbeta}= 
\frac{\exp(\beta_0 + \sum_{j=1}^p \beta_j x_{ij})}
{ 1+\exp(\beta_0 + \sum_{j=1}^p \beta_j x_{ij}) }, 
\quad i=1,\ldots,n,
\end{equation}
where $x_{i1},\ldots, x_{ip}$ are the covariates for individual $i$ and $\beta_0,\beta_1,\ldots,\beta_p$  unknown parameters. We adopt a probabilistic framework by assuming that 
 $\bx_i = (x_{i1},\ldots, x_{ip})$ is normally distributed:
\begin{equation*}
\bx_i \iid \mathcal{N}_p(\mu,\Sigma), \quad i=1,\cdots,n. 
\end{equation*}
Let $\theta=(\mu, \Sigma, \bbeta)$ be the set of parameters of the model. Then, the log-likelihood for the complete data can be written as:
\begin{equation*}
\begin{split}
\llike(\theta;\bx,\by) & =\sum_{i=1}^n \llike(\theta;\bx_i,y_i) \\
&=\sum_{i=1}^n \Big( \log  (\dens(y_i|\bx_i;\bbeta))+\log (\dens(\bx_i;\mu,\Sigma)) \Big).
\end{split}
\end{equation*}

Our main goal is to estimate the vector of parameters $\bbeta=(\beta_j,0\leq j \leq p)$ 
 when missing values exist in the design matrix, i.e., in the matrix $\bx$.
For each individual $i$, we note $\xiobs$ the elements of $\bx_{i}$ that are observed and $\ximis$ those that are missing. We also decompose the matrix of covariates  as $\bx = (\xobs,\xmis)$, keeping in mind that the missing elements may differ from one individual to another.

For each individual $i$, we define the missing data indicator vector $M_i=(M_{ij}, 1 \leq j \leq p)$, with $M_{ij}=1$ if $x_{ij}$ is missing and $M_{ij}=0$ otherwise. The matrix $M=(M_i, 1\leq i \leq n)$ then defines the missing data pattern. The missing data mechanism is characterized by the conditional distribution of $M$ given $x$ and $y$, with parameter $\phi$, i.e.,
$\dens(M_i|\bx_i,y_i,\phi).$ Throughout this paper, we assume a missing at random (MAR) mechanism which implies that the missing values mechanism  can therefore be ignored \citep{little_rubin} and  the maximum likelihood estimate of $\theta$ can be obtained by maximizing $\llike(\theta ; y,\xobs)$. A reminder of these concepts is given in $\ref{ann:ignorable}$.

\section{Parameter estimation by SAEM}\label{sectionsaem}
\subsection{The EM and MCEM algorithms}
We aim to estimate the parameter $\theta$ of the logistic regression model by maximizing the observed log-likelihood $\llike(\theta;\xobs, \by)$.
Let us start with the classical EM formulation for obtaining the maximum likelihood estimator from incomplete data. 
Given some initial value $\theta_0$, iteration $k$ updates $\theta_{k-1}$ to $\theta_k$ with the following two steps:
\begin{itemize}
\item \textbf{E-step:} Evaluate the quantity
\begin{equation}\label{eq::estep}
\begin{split}
Q_{k}(\theta)&=\mathbb{E}[\llike(\theta;\bx,\by)|\xobs,\by;\theta_{k-1}]\\
&=\int \llike(\theta;\bx,\by) \dens(\xmis|\xobs,\by;\theta_{k-1})d\xmis.
\end{split}
\end{equation}
\item \textbf{M-step:} Update the estimation of $\theta$:
$\theta_k = \argmax_{\theta}Q_{k}(\theta).$
\end{itemize}
Since the expectation (\ref{eq::estep}) in the E-step for the logistic regression model has no explicit expression, MCEM  \citep{mcemWeiTanner,ibrahim1999_MonteCarlo} can be used.  The E-step of MCEM generates several samples of missing data from the target distribution $\dens(\xmis|\xobs,\by;\theta_{k-1})$ and replaces the expectation of the complete log-likelihood by an empirical mean. However, an accurate Monte Carlo approximation of  the E-step may require a significant computational effort, as illustrated in the Section \ref{sectionsimu}.

\subsection{The SAEM algorithm}
To achieve improved computational efficiency, we suggest deriving a SAEM algorithm \citep{lavielle:hal-01122873} which replaces the E-step (\ref{eq::estep}) by a stochastic approximation. 
Starting from an initial guess $\theta_0$, the $k$th iteration consists of three steps:
\begin{itemize}
\item \textbf{Simulation:} For $i=1,2,\cdots,n$, draw $\ximis^{(k)}$ from 
\begin{eqnarray}\label{eq::target}
\dens(\ximis|\xiobs,\by_i;\theta_{k-1}).
\end{eqnarray}

\item \textbf{Stochastic approximation:} Update the  function $Q$ according to 
\begin{equation}\label{eq::sa}
Q_{k}(\theta)=Q_{k-1}(\theta)+\gamma_k\left(\llike(\theta ;\xobs,\xmis^{(k)},\by)-Q_{k-1}(\theta)\right),
\end{equation}
where $(\gamma_k)$ is a non-increasing sequence of positive number.
\item \textbf{Maximization:} Update the estimation of $\theta$:
\begin{equation*}
\theta_k = \argmax_{\theta}Q_{k}(\theta).
\end{equation*}
\end{itemize}
The choice of the sequence $(\gamma_k)$ in (\ref{eq::sa}) is important for ensuring the almost sure convergence of SAEM to a maximum of the observed likelihood \citep{convergence_saem}.
We will see in Section \ref{sectionsimu} that, in our case, very good convergence is obtained using $\gamma_k=1$ during the first iterations, followed by a sequence that decreases as $1/k$.

\subsection{Metropolis-Hastings sampling}

In the logistic regression case, the unobserved data cannot in general be drawn exactly from the conditional distribution \eqref{eq::target}, which depends on an integral that is not calculable in closed form. One solution is to use a Metropolis-Hastings (MH) algorithm, which consists of constructing a Markov chain that has the target distribution as its stationary distribution. The states of the chain after $S$ iterations are then used as a sample from the target distribution. 
To define a proposal distribution for MH algorithm, we observe that the target distribution \eqref{eq::target} can be factorized as follows:
$$\dens(\ximis|\xiobs,\by_i;\theta)\propto \dens(y_{i}|\bx_{i};\bbeta)\dens(\ximis|\xiobs;\mu,\Sigma). $$
We select the proposal distribution as the second term $\dens(\ximis|\xiobs,\mu,\Sigma)$, which is normally distributed:
\begin{equation}\label{functiong}
\ximis|\xiobs  \sim \mathcal{N}_p(\mu_i,\Sigma_i),
\end{equation}
where
\begin{equation*}
\begin{split}
\mu_i&=\mu_{i, {\rm mis}}+\Sigma_{i,{\rm mis,obs}}\Sigma_{i,{\rm obs,obs}}^{-1}(\xiobs-\mu_{i, {\rm obs}}),\\
\Sigma_i&=\Sigma_{i,{\rm mis,mis}} -\Sigma_{i,{\rm mis,obs}} \Sigma_{i,{\rm obs,obs}}^{-1}\Sigma_{i,{\rm obs,mis}},
\end{split}
\end{equation*}
with $\mu_{i, {\rm mis}}$ (resp.  $\mu_{i, {\rm obs}}$)  the missing (resp. observed) elements of $\mu$ for individual $i$.  The covariance matrix $\Sigma$ is decomposed in the same way.
The MH algorithm is described further in \ref{ann:mh}. 


\subsection{Observed Fisher information}\label{sectionfish}
After computing the MLE $\thml$ with SAEM, we estimate its variance. To do so, we can use the observed Fisher information matrix (FIM): $
\mathcal{I}(\theta) = -\frac{\partial^2 \llike(\theta;\xobs,\by)}{\partial \theta \partial \theta^T}. $
According to Louis' formula \citep{louis}, we have:
\begin{equation*}
\begin{split}
\mathcal{I}(\theta) = &-\mathbb{E}\left(\frac{\partial^2 \llike(\theta;\bx,\by)}{\partial \theta \partial \theta^T}\big|\xobs,\by;\theta \right) \\
&- \mathbb{E}\left(\frac{\partial \llike(\theta;\bx,\by)}{\partial \theta} \frac{\partial \llike(\theta;\bx,\by)^T}{\partial \theta}\big|\xobs,\by;\theta \right)\\
&+\mathbb{E}\left(\frac{\partial \llike(\theta;\bx,\by)}{\partial \theta} |\xobs,\by;\theta \right)\mathbb{E}\left(\frac{\partial \llike(\theta;\bx,\by)}{\partial \theta} |\xobs,\by;\theta \right)^T.
\end{split}
\end{equation*}
The observed FIM can therefore be expressed in terms of conditional expectations, which can also be approximated using a Monte Carlo procedure. More precisely, 
given $S$ samples $(\ximis^{(s)}, 1\leq i \leq n, 1 \leq s \leq S)$ of the missing data drawn from the conditional distribution (\ref{eq::target}), the observed FIM can be estimated as
$\hat{\mathcal{I}}_S(\hat{\theta})= \sum_{i=1}^n -(D_{i}+G_i-\Delta_i \Delta_i^T),$
where 
\begin{equation*}
\begin{split}
\Delta_i &= \frac{1}{S}\sum_{s=1}^S \frac{\partial \llike(\hat{\theta};\ximis^{(s)},\xiobs,y_i)}{\partial \theta},\\
D_i &= \frac{1}{S}\sum_{s=1}^S \frac{\partial^2 \llike(\hat{\theta}; \ximis^{(s)},\xiobs,y_i)}{\partial \theta\partial \theta^T},\\
G_i &= \frac{1}{S}\sum_{s=1}^S \left(\frac{\partial \llike(\hat{\theta};\ximis^{(s)},\xiobs,y_i)}{\partial \theta}\right) \left( \frac{\partial \llike(\hat{\theta};\ximis^{(s)},\xiobs,y_i)}{\partial \theta} \right)^T.\\
\end{split}
\end{equation*}
Here, the gradient and the Hessian matrix can be computed in closed form. 
The procedure for calculating the observed information matrix is described in  \ref{ann:var}.

\section{Model selection and prediction}\label{sectionmodselect}

\subsection{Information criteria}

In order to compare different possible covariate models, we can consider penalized likelihood  criteria such as the Bayesian information criterion (BIC). For a given model ${\cal M}$ and an estimated parameter $\hat{\theta}_{\cal M}$, BIC is defined as:
\begin{equation*}
\begin{split}
{\rm BIC}({\cal M}) &= -2\llike(\hat{\theta}_{\cal M}; \xobs,\by) + 
\log(n)d({\cal M}), 
\end{split}
\end{equation*}
where $d({\cal M})$ is the number of estimated parameters in a model ${\cal M}$. 
The distribution of the complete set of covariates 
$(x_{ij}, 1\leq i \leq n, 1 \leq j \leq p)$ does not depend on the regression model used for  modeling the binary outcomes $(y_i, 1 \leq i \leq n)$: we assume the same normal distribution ${\cal N}_p(\mu,\Sigma)$ for all regression models. 
Thus, the difference between models between the number $d({\cal M})$  of estimated parameters is equivalent to  the difference between the number of non-zero coefficients in $\beta_{\cal M}$. Note that, contrary to the suggested approach, the existing methods \citet{missaic,missaic2} use an approximation of the Akaike information criterion (AIC) without estimating the observed likelihood.
\subsection{Observed log-likelihood}\label{subsectionoll}
For a given model and parameter $\theta$, the observed log-likelihood
is, by definition:
$$\llike({\theta}; \xobs,\by) = \sum_{i=1}^n \log \left(\dens(y_i,\xiobs;\theta) \right).$$
With missing data, the density $\dens(y_i,\xiobs;\theta)$ cannot in general be computed in closed form. We suggest to approximate it using an importance sampling Monte Carlo approach. Let $g_i$ be the density function of the normal distribution defined in \eqref{functiong}. Then,
\begin{equation*}
\begin{split}
\dens(y_i,\xiobs;\theta) 
&= \int \dens(y_i,\xiobs|\ximis;\theta)\dens(\ximis;\theta) d\ximis  \\
&= \int \dens(y_i,\xiobs|\ximis;\theta)\frac{\dens(\ximis;\theta)}{g_i(\ximis)} g_i(\ximis) d\ximis  \\
&= \mathbb{E}_{g_i} \left(\dens(y_i,\xiobs|\ximis;\theta)\frac{\dens(\ximis;\theta)}{g_i(\ximis)}\right). 
\end{split}
\end{equation*}
Consequently, if we draw $M$ samples  from the proposal distribution  (\ref{functiong}):  
$$\ximis^{(s)} \iid {\cal N}(\mu_i,\Sigma_i), \quad m=1,2,\cdots,S,$$
we can estimate $\dens(y_i,\xiobs;\theta)$ by:
$$\hat\dens(y_i,\xiobs;\theta) = \frac{1}{S}\sum_{m=1}^S \dens(y_i,\xiobs|\ximis^{(s)};\theta)\frac{\dens(\ximis^{(s)};\theta)}{g_i(\ximis^{(s)})},$$
and derive an estimate of the observed log-likelihood $\llike({\theta}; \xobs,\by)$.

\subsection{
Prediction on test set with missing values}
\label{sectionpredna}
In supervised learning, after fitting a model using a training set, a natural step is to evaluate the prediction performance, which can be done with a test set. Assuming $\bx=(\xobs,\xmis)$ an observation in the test set, we want to predict the binary response $y$. One important point is that test set also contains missing values, since the training set and the test set have the same distribution (\textit{i.e.}, the distribution of covariates and the distribution of missingness). Therefore, we can't directly apply  the fitted model  (which uses $p$ coefficients)  to predict $y$ from an incomplete observation of the test $\bx$.\\

Our framework offers a natural way to tackle this issue by marginalizing over the distribution of missing data  given the observed ones.
More precisely, with $S$ Monte Carlo samples $$(\xmis^{(s)}, 1 \leq s \leq S) \sim \dens(\xmis|\xobs),$$
we estimate directly the response by maximum a posteriori
\begin{equation*}
\begin{split}
\hat{y} = \argmax_{y} \dens(y|\xobs) 
&= \argmax_{y} \int \dens(y|\bx) \dens(\xmis|\xobs) d \xmis\\
&= \argmax_{y} \mathbb{E}_{\dens_{\xmis|\xobs}} \dens(y|\bx)\\
&= \argmax_{y} \sum_{s=1}^S \dens \left(y|\xobs,\xmis^{(s)} \right).
\end{split}
\end{equation*}
Note that in the literature there are not many solutions to deal with the missing values in the test set. 
In Subsection \ref{subsection:pred}, we compare the suggested approach to some methods used in practice based on imputation of the test set.

\section{Simulation study}\label{sectionsimu}

\subsection{Simulation settings}\label{subsection:simu}

We first generated a design matrix $\bx$ of size $n=1000$ $\times$ $p=5$ by drawing each observation from a multivariate normal distribution $\mathcal{N}(\mu, \Sigma)$. Then, we generated the response according to the logistic regression model \eqref{regmodel}.
We considered as the true parameter values:
$\bbeta = (-0.2, 0.5, -0.3, 1, 0, -0.6)$,
$\mu = (1,2,3,4,5)$, 
$\Sigma = \text{diag}(\sigma)C \text{diag}(\sigma$),
where the $\sigma$ is the vector of standard deviations 
$\sigma=(1,2,3,4,5)$,  
and  $C$ the correlation matrix 
\begin{equation}\label{eq:C}
C = \begin{bmatrix}
1  & 0.8 & 0 & 0 &   0\\
0.8 & 1 & 0 & 0  &  0\\
0  & 0 & 1 & 0.3 &   0.6\\
0 & 0 & 0.3 & 1 &  0.7\\
0 & 0 & 0.6 & 0.7 &  1\\
\end{bmatrix}\,.
\end{equation}

Before generating missing values, we performed classical logistic regression on the complete dataset, the results (ROC curve) are provided in \ref{ann:orig}.
Then we randomly introduced 10\% missing values in the covariates first with a missing completely at random (MCAR) mechanism where each entry has the same probability to be observed.
\begin{figure}[!htbp]
\centering
 \makebox{\includegraphics[width=0.7\textwidth]{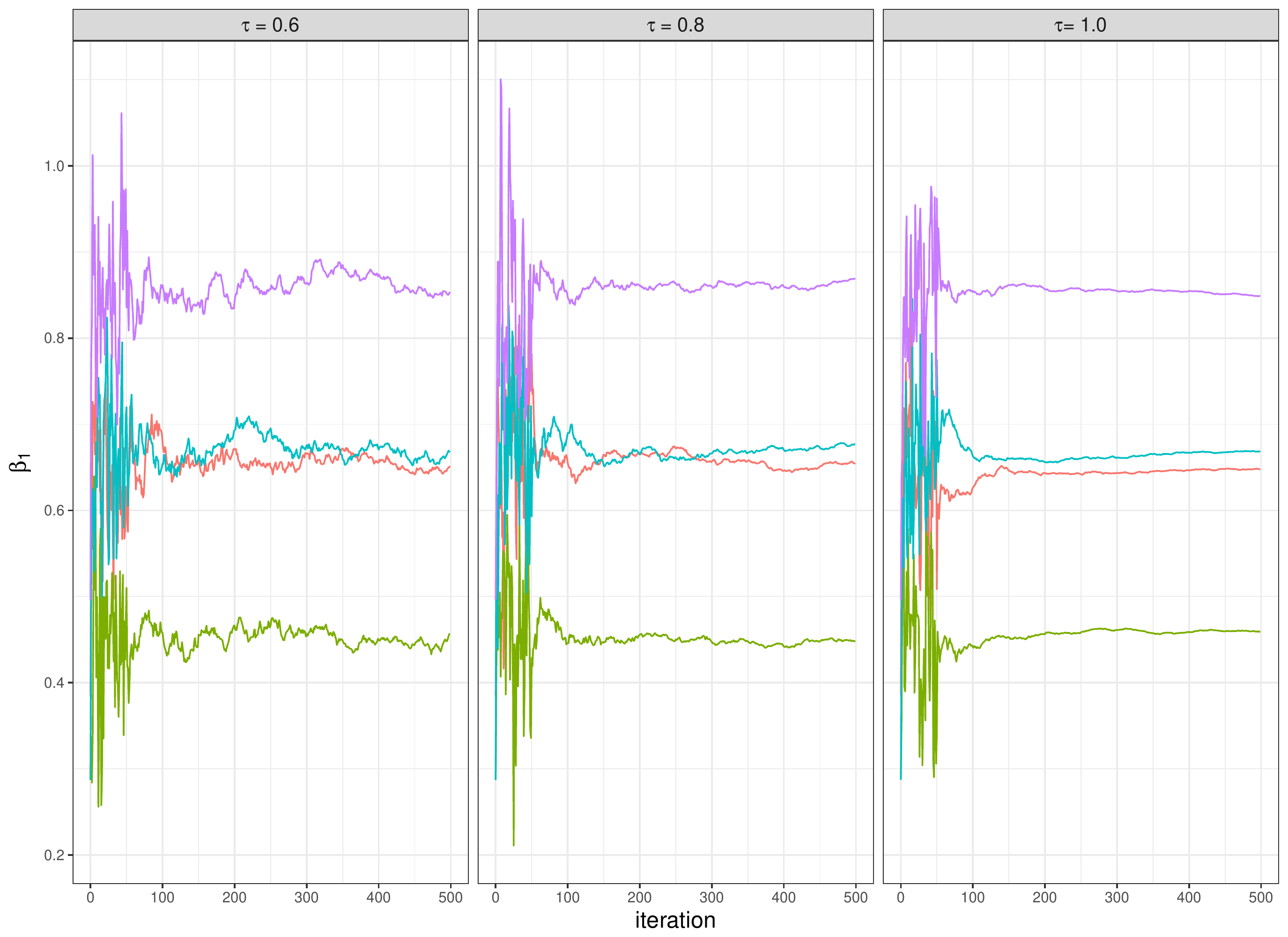}}
\caption{\label{fig:convergence}Convergence plots for $\beta_1$ obtained with three different values of $\tau$ (0.6, 0.8, 1.0). Each color represents one simulation. The true value of $\beta_1 = 0.5.$}
\end{figure}
\subsection{The behavior of SAEM}\label{subsect::saem}
The algorithm was initialized with the parameters 
obtained after mean imputation, i.e., where missing value in a variable are replaced by the unconditional mean calculated from the the available cases and the logistic regression is applied on the completed data. For the non-increasing sequence $(\gamma_k)$ in the Stochastic Approximation step of SAEM,
we chose $\gamma_k = 1$ during the first $k_1$ iterations in order to converge quickly to a neighborhood of the MLE, and from $k_1$ iterations on, we set $\gamma_k = (k - k_1)^{-\tau}$ to assist the almost sure convergence of SAEM.
In order to study the effect of the sequence of stepsizes $(\gamma_k)$,  
we fixed the value of $k_1=50$ and used $\tau=(0.6 , \ 0.8, \ 1)$ during the next 450 iterations. 
Representative plots of the convergence of SAEM for the coefficient $\beta_1$, obtained from four simulated data sets, are shown in Figure \ref{fig:convergence}.
For larger $\tau$, 
 SAEM  converged faster, and with less fluctuation. 
For a given simulation, the three sequences of estimates converged to the same solution, but using $\tau=1$ yielded the fastest convergence, and showed less fluctuation.
We therefore use $\tau=1$ in the following.

\begin{figure}[!htbp]
\centering
    \makebox{
\includegraphics[width=0.8\textwidth]{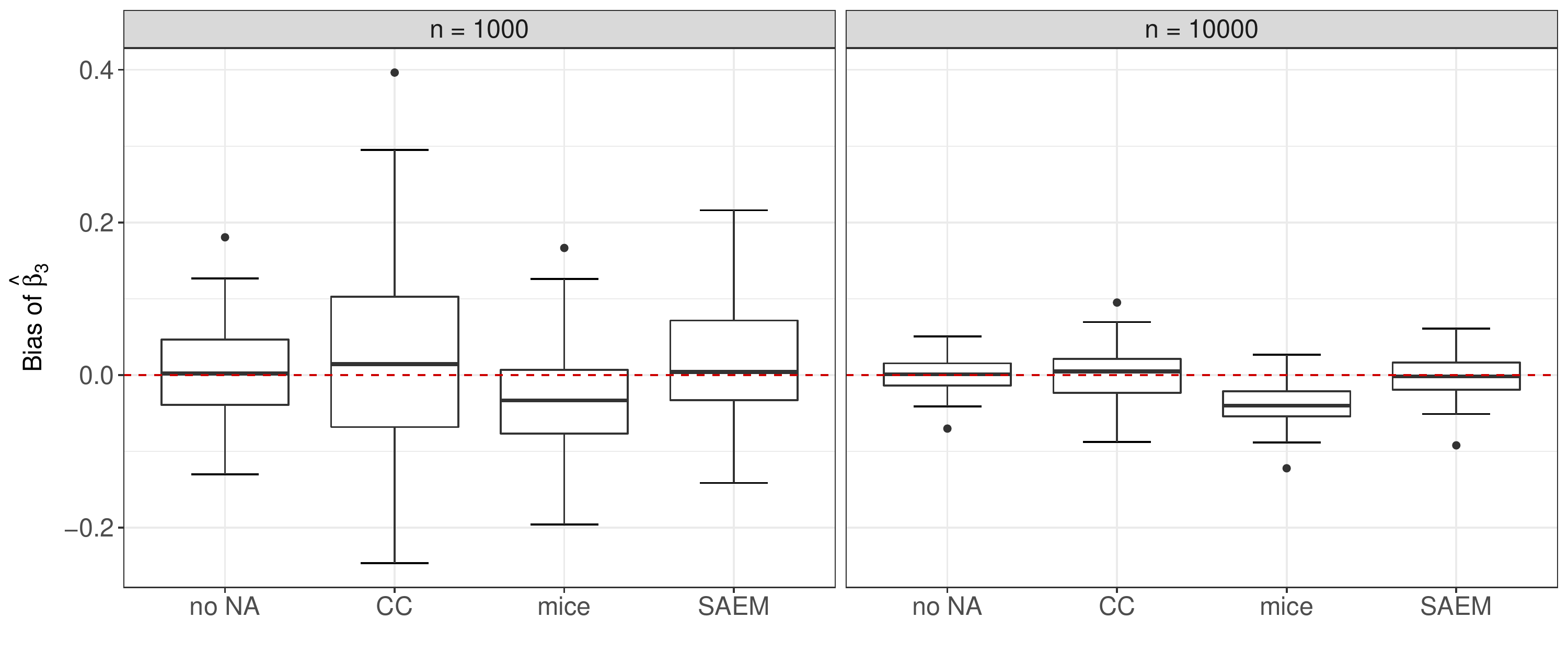}}
 \makebox{
\includegraphics[width=0.8\textwidth]{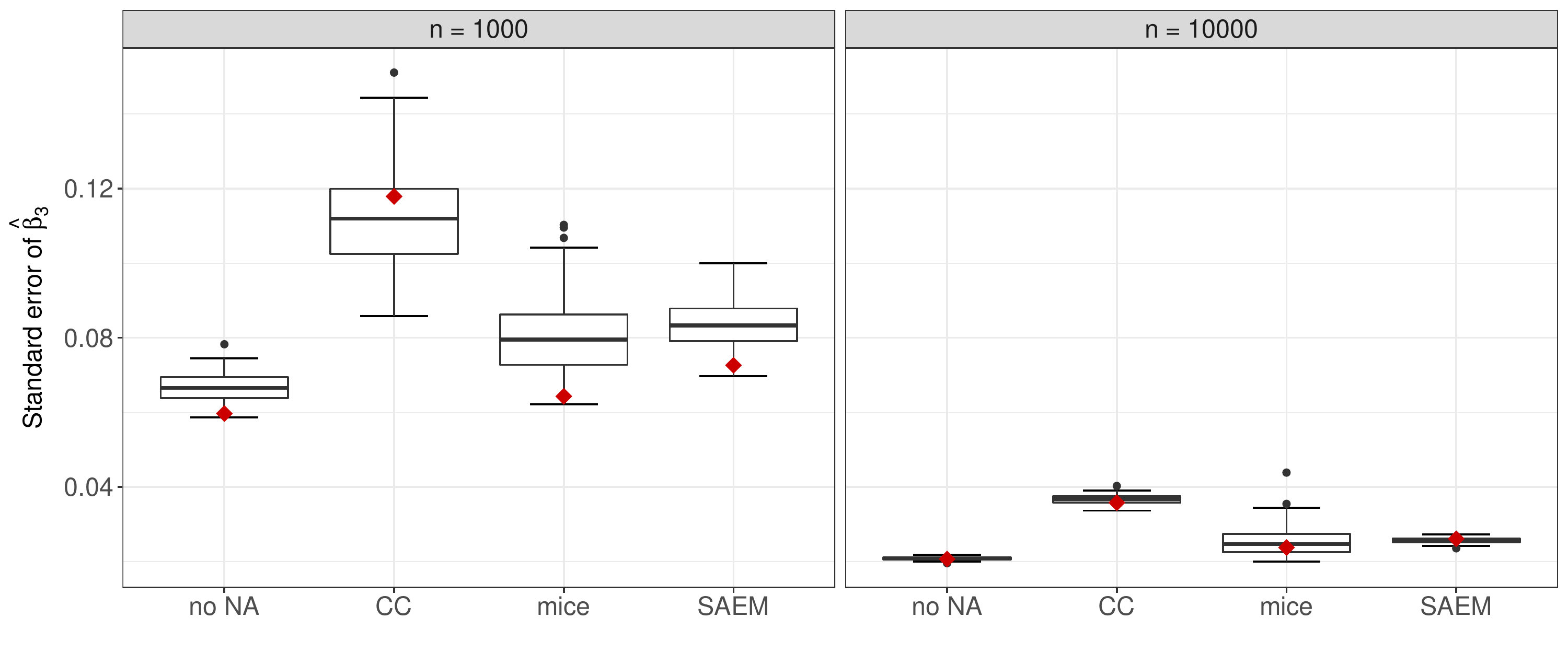}
}
\caption{\label{n_bias} Top: Empirical distribution of bias of $\hat\beta_3$.
Bottom: Distribution of the estimated standard errors of $\hat\beta_3$; for each method,  the red point corresponds to the empirical standard deviation of $\hat{\beta}_3$ calculated over the 1000 simulations.
Results for 10\% MCAR and correlation $C$.}
\end{figure} 
\subsection{Comparison with other methods}
\label{subsect::simucomp}
We ran $1000$ simulations and compared SAEM to several other existing methods, initially in terms of estimation errors of the parameters. 
We mainly focused on \emph{i)} the complete case (CC) method, i.e., all rows containing at least one unobserved data value were removed, \emph{ii)} multiple imputation by chained equations (mice) with Rubin's combining rules 
\citep{mice}. More precisely, missing values are imputed successively by drawing from conditional distribution. We use the default arguments of the function implemented in R, i.e., conditional models based on regression models are used for quantitative variables and on logistic regression models are used for binary variables and uncertainty of the parameters is reflected within a Bayesian framework. More details are in \citet{mice}.
Finally, we used the dataset without missing values (no NA) as a reference, with parameters estimated with the Newton-Raphson algorithm. 
We varied the number of observations $n=200, 1000$ and $10\,000$, the missing value mechanism MCAR and MAR, the percentage of missing values $10\%$ and $30\%$, as well as the correlation structure either using $C$ given by (\ref{eq:C}) or an orthogonal design.

Figure \ref{n_bias} (top) displays the distribution of the estimates of $\beta_3$, for $n=1000$ and $n=10\,000$ under MCAR mechanism and the correlation between covariates is given by (\ref{eq:C}). Results of simulation with $n=200$ are presented in supplementary materials \citep{supp}.
This plot is representative of the results obtained  with the other components of $\beta$. 
As expected, larger samples yielded less variability. 
Moreover,  we observe that in both cases, the estimation obtained by mice could be biased, whereas SAEM provided unbiased estimates with small variances.
Figure \ref{n_bias} (bottom) represents the empirical distribution of the estimated standard error of $\hat\beta_3$. For SAEM it was calculated using the observed Fisher information as described in Section \ref{sectionfish}. 
With a larger $n$, not only the estimated standard errors, but also variance of estimation, clearly decreased for all of the methods. In the case where $n=1000$, SAEM and mice slightly overestimated the standard error, while CC underestimated it, on average. 
Globally, SAEM led to the best result, since compared with its competitor mice, it had a similar estimation of the standard error on average, but with much less variance. 

\begin{table}[!htbp]
\caption{\label{coverage} Coverage (\%) for $n=10\,000$, correlation $C$ and $10\%$ MCAR, calculated over 1000 simulations. Bold indicates under coverage. Inside the parentheses is the average length of corresponding confidence interval over 1000 simulations (multiplied by 100).} 
\centering
\fbox{
\begin{tabular}{lcccr}%
  parameter & no NA & CC & mice & SAEM \\ 
  \hline
  $\beta_0$ & 95.2 (21.36) & 94.4 (27.82) & 95.2 (22.70) & 94.9 (22.48)\\
    $\beta_1$ & 96.0 (18.92) & 94.7 (24.65) & 93.9 (21.77) & 95.1 (21.51) \\
      $\beta_2$ & 95.5 (9.53) & 94.6 (12.41) & 94.0 (10.97) & 94.3 (10.83)\\
        $\beta_3$ & 94.9 (8.17) & 94.3 (10.66) & \textbf{86.5} (9.03) & 94.7 (9.03)\\
          $\beta_4$ & 94.6 (4.00) & 94.2 (5.21) & 96.2 (4.49) & 95.4 (4.42)\\
            $\beta_5$ & 95.9 (5.52) & 94.4 (7.19) & \textbf{89.6} (6.20) & 94.7 (6.17)
\end{tabular}
}
\end{table}
Table \ref{coverage} shows the coverage of the confidence interval for all parameters and inside the parentheses is the average length of corresponding confidence interval. We had expected coverage at the nominal 95\% level. The simulation margin of error corresponding to coverage results is 1.35\%. SAEM reached from 94.3\% to 95.4\%  coverage, while mice struggled for certain parameters: the coverage rates for a few estimates are  $89.6\%$ or $86.5\%$, which are significantly below the nominal level.  Even though CC showed reasonable results in terms of coverage, the width of its confidence interval was still too large. Simulation with smaller sample size had the same results, for example, coverages for $n=200$ are presented in  supplementary materials \citep{supp}.

\begin{table}[!htbp]
\caption{\label{compmicrobenchmark} Comparison of execution time between no NA, MCEM, mice, and SAEM with correlation $C$ and $10\%$ MCAR, for $n=200$ or $n=1000$, calculated over 1000 simulations. }
\centering
\fbox{%
\begin{tabular}{lcccr}
  Execution time (seconds) \\ for one simulation &no NA& MCEM & mice & SAEM \\ 
  \hline
  \multicolumn{5}{l}{$\bm{n=1000}$}\\
min  &$2.87\times 10^{-3}$ & 492 & 0.64 & 9.96 \\
mean &$4.65\times 10^{-3}$ & 773 & 0.70 & 13.50 \\
max  &$43.50\times 10^{-3}$ &1077 & 0.76 & 16.79\\ 
  \hline
\multicolumn{5}{l}{$\bm{n=200}$}\\
  min  &$1.26\times 10^{-3}$ & 67.91 & 0.24 & 2.64 \\
  mean &$2.32\times 10^{-3}$ & 291.47 & 0.28 & 3.91 \\
  max  &$21.53\times 10^{-3}$ &1003 & 0.48 & 6.04 
\end{tabular}}
\end{table}
  Lastly, Table \ref{compmicrobenchmark} highlights large differences between the methods in terms of execution time.  In fact we also implemented MCEM algorithm \citep{ibrahim1999_MonteCarlo} using adaptive rejection sampling. Even with a very small sample size $n=200$, MCEM took on average 5 minutes for one simulation; while multiple imputation took less than 1 second per simulation, and  SAEM less than 10 seconds, which remains reasonable. However, the bias and standard error for the estimation of SAEM and MCEM were quite similar, as presented in supplementary materials \citep{supp}. Due to this computational difficulty, we didn't perform MCEM to compare with others in the experiments with larger sample sizes.

\begin{figure}[!htbp]
\centering
\makebox{\includegraphics[width=0.8\textwidth]{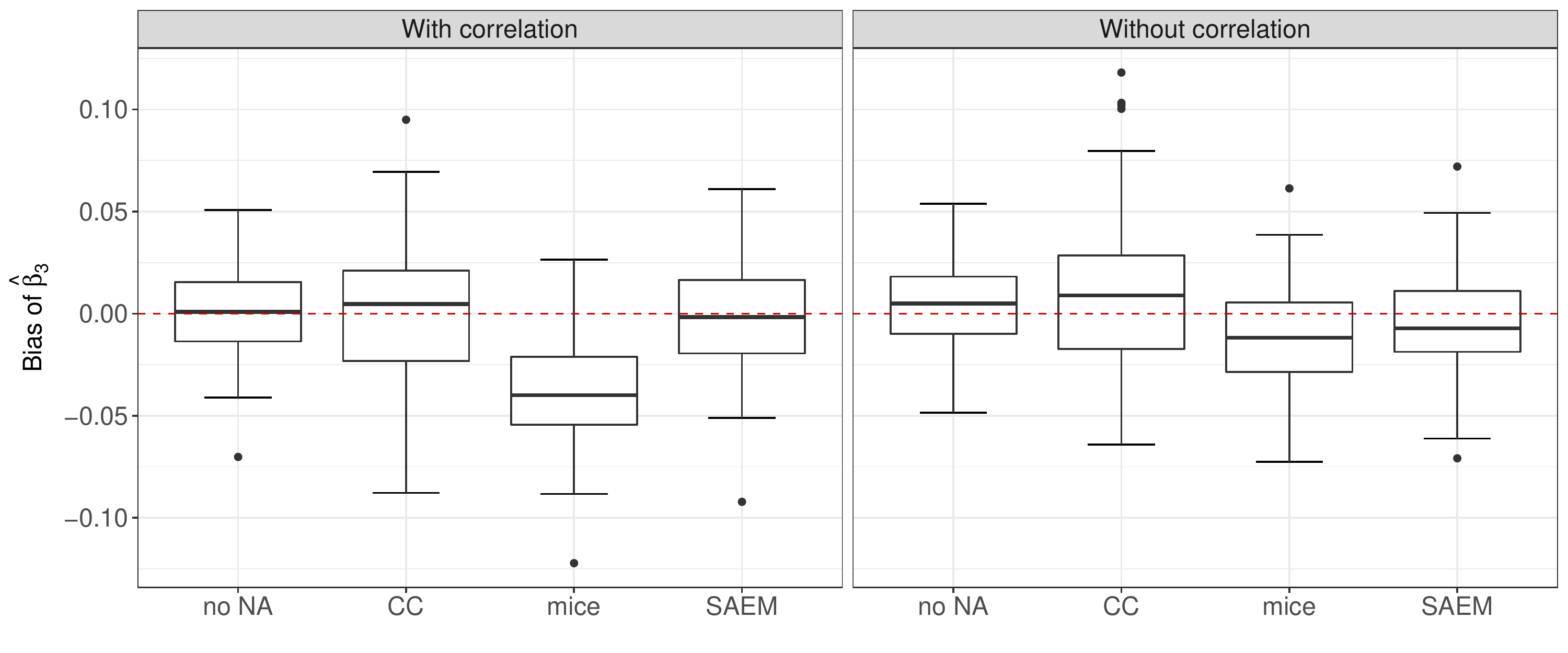}}
\caption{\label{fig:cor} Empirical distribution of the estimates of $\beta_3$ obtained  under MCAR, with $n=10\,000$ and 10\% of missing values; left: the covariates are correlated; right: no correlation between the covariates.}
\end{figure} 
The results obtained, when the covariates were independent, are also presented. Figure \ref{fig:cor} (right) shows the results of estimation in the case with orthogonal design. SAEM was a little biased since it estimated non-zero terms for the covariance, but it stills outperformed CC and mice.

\subsection{Extended simulations}

\paragraph{Missing at Random mechanisms.}
We first simulated a binary vector $\eta = (\eta_1, \eta_2, \cdots, \eta_p)$ of dimension $n \times p$ from Bernoulli distribution, where $\eta_{ij} = 0$ indicates that the corresponding  $x_{ij}$ will be missing while $1$ indicates observed. Then the probability of having missing data on one variable is calculated by a logistic regression function. For example in our case $p=5$ and the realizations of $\eta$ (the pattern) $(1,0,1,0,0)$, the probability that covariates $(x_2, x_4, x_5)$ can be missing, depends only on $x_1$ and $x_3$ with a logistic regression model. The weights in the linear combination impact the proportion of missingness. 
We introduced 10\% of missing values in the covariates according to the MAR mechanisms. 
The results presented in \ref{ann:mar} highlight that as expected they are similar to the ones obtained under MCAR and the parameters are estimated without bias. 

\paragraph{Robustness to the Gaussian assumption for covariates.}

First we generated a design matrix of size $n=1000$ $\times$ $p=5$ by drawing each observation from a multivariate Student distribution $t_v(\mu, \Sigma)$ with degree of freedom $v=5$ or $v=20$, and $(\mu, \Sigma)$ the same as those in Normal distribution in Subsection \ref{subsection:simu}.  Then, we considered the Gaussian mixture model case by generating half of the samples from $\mathcal{N}(\mu_1, \Sigma)$ and the other half from $\mathcal{N}(\mu_2, \Sigma)$, where $\mu_1 = (1,2,3,4,5)$ and $\mu_2=(1,1,1,1,1)$, and the same  $\Sigma$ as previously. Then, we generated the response according to the same logistic regression model as described in Subsection \ref{subsection:simu} and considered either MCAR or MAR mechanisms.
\begin{figure}[!htbp]
\centering
\makebox{\includegraphics[width=0.75\textwidth]{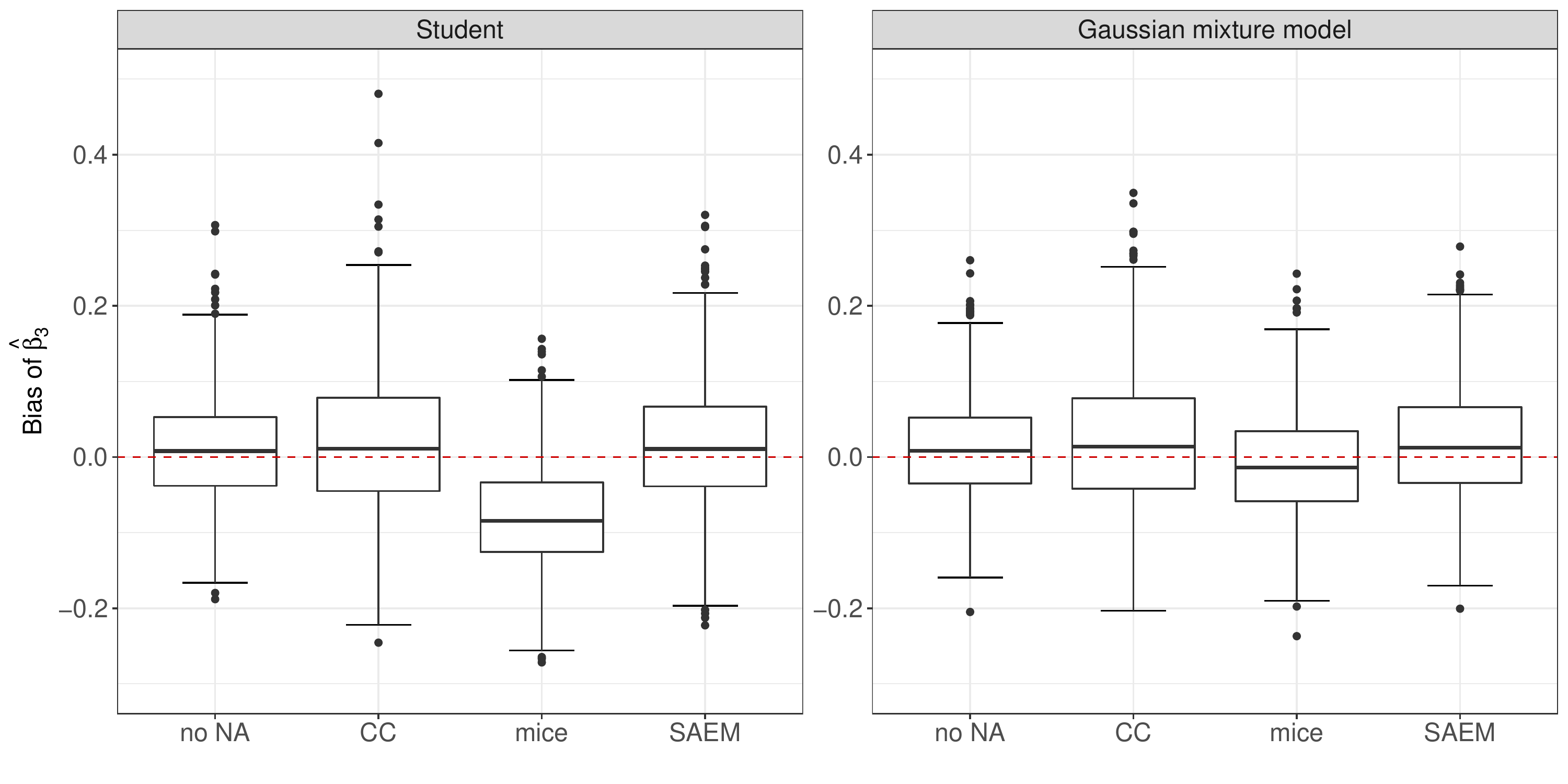}}
\caption{\label{fig:st} Empirical distribution of the bias of $\hat{\beta}_3$ obtained for misspecificated models under MCAR, with $n=1000$; left: Student distribution with degree of freedom $v=5$; right: Gaussian mixture model.}
\end{figure} 

Figure \ref{fig:st} illustrates the estimation bias of the parameter $\beta_3$ and  \ref{ann:miscoverage} shows the coverage for all parameters and inside the parentheses is the average length of corresponding confidence interval. 
This experiment shows that the estimation bias for regression coefficient with the proposed method even based on normal assumption, is robust to such a model misspecification. Indeed, the bias may increase when covariates don't follow exactly a normal distribution, but the increase is negligible compared to the bias of imputation based methods. We also observe only a small undercoverage compared to mice, and a more reasonable length of confidence interval compared to CC. 

\paragraph{Varying the percentage of missing values}

When the percentage of missing values increases, the variability of the results increases but the methods still provide satisfactory results as illustrated in supplementary materials \citep{supp}. 

\paragraph{Varying the separability of the classes}

When the classes are very separated SAEM can exibit a biais and large variance as illustrated in supplementary materials \citep{supp}. However, the logistic regression without missing values also encounters difficulties. 

In summary, not only did these simulations allow us to verify that SAEM  leads to estimators with limited bias, but also they ensured that we made correct inferences by taking into account the additional variance due to missing data.

\subsection{Model selection}

To look at the capabilities of the method in terms of model selection, we considered the same simulation scenarios as in Section \ref{subsection:simu}, with some parameters set to zero. 
We now describe the results for the case where all parameters in $\beta$ are zero except $\beta_0=-0.2$, $\beta_1=0.5$, $\beta_3=1$ and $\beta_5=-0.6$. 
We compared the $BIC_{obs}$ based on the observed log-likelihood, as described in Section \ref{sectionmodselect}, to that based on the complete cases $BIC_{cc}$ and that obtained from the the original complete data $BIC_{orig}$. 

\begin{table}[!htbp]
\caption{\label{ms1} For data with or without correlations, the percentage of times that each criterion selects the correct true model (C), overfits (O), and underfits (U).}
\centering
\fbox{%
\begin{tabular}{l c c c @{\hskip 0.5in} c c r }
& \multicolumn{3}{l}{Non-Correlated} & \multicolumn{3}{l}{Correlated}\\
  Criterion & C & O & U  & C & O & U \\
  \hline
  $BIC_{obs}$ &  92&  3& 5&  94&  2& 4\\
  $BIC_{orig}$ &  96&  2& 2&  93&  0& 7\\
  $BIC_{cc}$ &  79&  1& 20&  91&  0& 9\\
\end{tabular}}
\end{table}
Table \ref{ms1} shows, with or without correlation between covariates, the percentage of cases where  each criterion selects the true model (C), overfits (O) -- i.e., selects more variables than there were -- or underfits (U) -- i.e., selects less variables than there were. In the case where the variables were correlated, the correlation matrix was the same as in  Section \ref{subsection:simu}.
 These results are representative of those obtained with other simulation schemes.


\subsection{\textcolor{black}{Prediction on a test set with missing values}}
\textcolor{black}{To evaluate the prediction performance on a test set with missing values, we considered the the same simulation scenarios for the training set as in Subsection \ref{subsection:simu} with sample size $1000 \times 5$. We also generated a test set of size $100 \times 5$. 
We compared the suggested approach described in Subsection \ref{sectionpredna}, with imputation methods. More precisely, we considered single imputation methods on the training set followed by classical logistic regression and variable selection by BIC on the imputed dataset such as
\emph{i)} imputation by the mean (impMean) \emph{ii)} imputation by PCA (impPCA) \citep{missMDA} which is based on low-rank assumption of the data matrix to impute. 
\emph{iii)} imputation by mice.  Note that \citet{hentges1998predictive} highlighted from a simulation study that, imputation methods can have good performance when the aim is to predict in logistic regression for MCAR data.
For all the imputation methods, we also imputed the test set independently and then applied the model that had been selected on the training set. Note that this can be a limitation if there is only one individual in the test set to predict whereas the suggested method does not encounter this issue.\\ 
We compared all these approaches with classical measures to evaluate predicted probability of logistic regression, such as AUC (area under the ROC curve), Brier score \citep{brier} and Logarithmic score  \citep{loga}. Figure \ref{fig:pred} shows that on average, marginalizing over distribution of missing values has the best performances: it gave the largest AUC and Logarithmic score, and the smallest Brier scores.
\begin{figure}[!htbp]
\centering
    \makebox{
\includegraphics[width=1\textwidth]{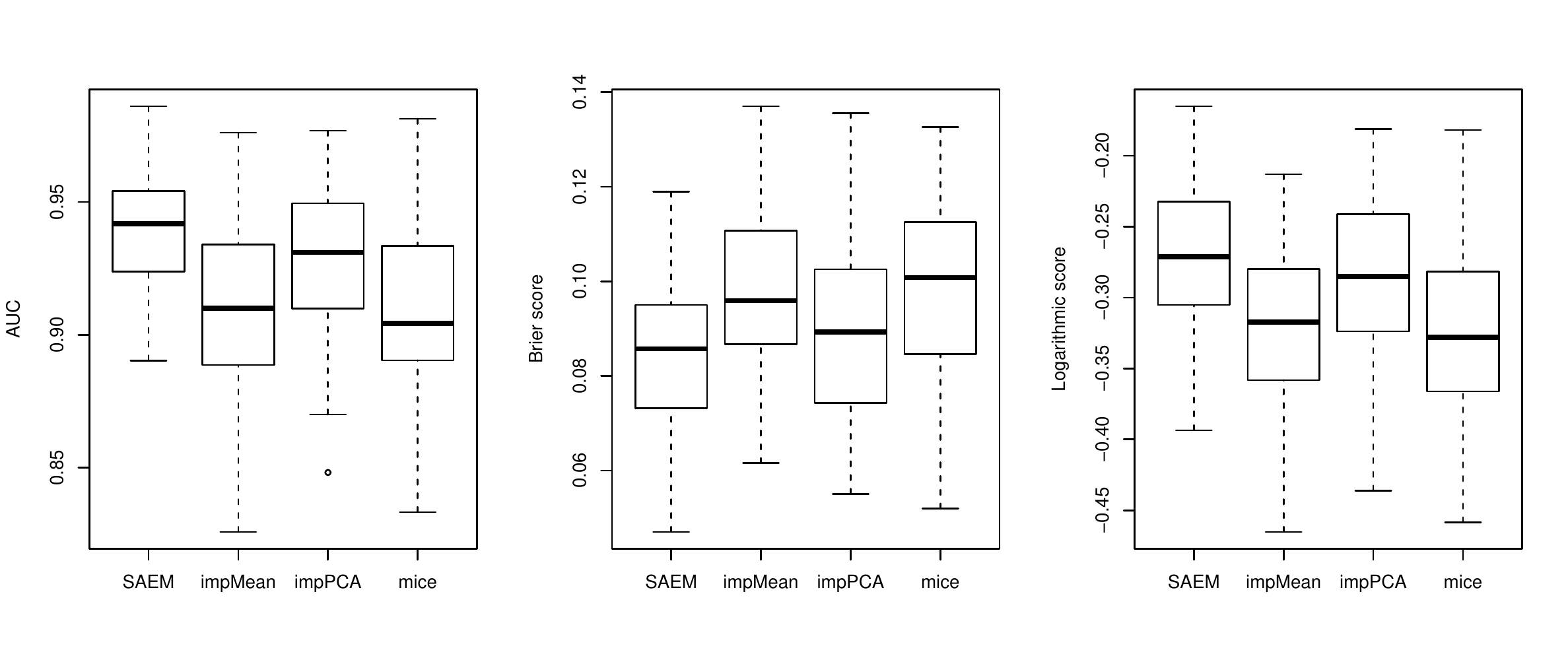}}
\caption{\label{fig:pred} Comparison of empirical distribution of AUC, Brier score and Logarithmic score obtained on the test set, for the proposed approach SAEM without imputation, impMean, impPCA and mice, over 100 simulations.}
\end{figure}
}

\section{Risk of severe hemorrhage for TraumaBase}\label{sectiontrauma}


The aim of our work is to accelerate and simplify the detection of patients presenting in hemorrhagic shock due to blunt trauma to speed up the management of this most preventable cause of death in major trauma. An optimized organization is essential to control blood loss as quickly as possible and to reduce mortality.

\subsection{Details on the dataset}\label{subsection:dataset}

This study has used the data collected from a trauma registry (TraumaBase$^{\tiny{\textregistered}}$) shared between six trauma centers within the Ile de France region (Paris area) in France. These centers have joined TraumaBase progressively between January 2011 and June 2015. Since then, data collection is exhaustive and covers the whole administrative area around Paris. The structure of the database integrates algorithm for consistency and coherence, and the data monitoring is performed by a central administrator. Sociodemographic, clinical, biological and therapeutic data (from the prehospital phase to the discharge if hospital) are systematically recorded for all trauma patients, and all patients transported in the trauma rooms of the participating centers are included in the registry.
As a result, there were 7495 individuals in the trauma data that we investigated, collected from January 2011 to March 2016, with age ranged from 12 to 96. The study group decided to focus on patients with blunt trauma to be able to compare to the existing prediction rules. Patients with pre-hospital cardiac arrest and missing pre-hospital data were excluded. 
After this selection,  6384 patients remained in the data set.
Based on clinical experience, 16 influential quantitative measurements were included. Detailed descriptions of these measurements and their histograms are  shown 
 in \ref{ann:trauma}. These variables were chosen because they were all available to the pre-hospital team, and therefore could be used in real situations.
 \begin{figure}[!htbp]
\centering
    \makebox{
\includegraphics[width=0.83\textwidth]{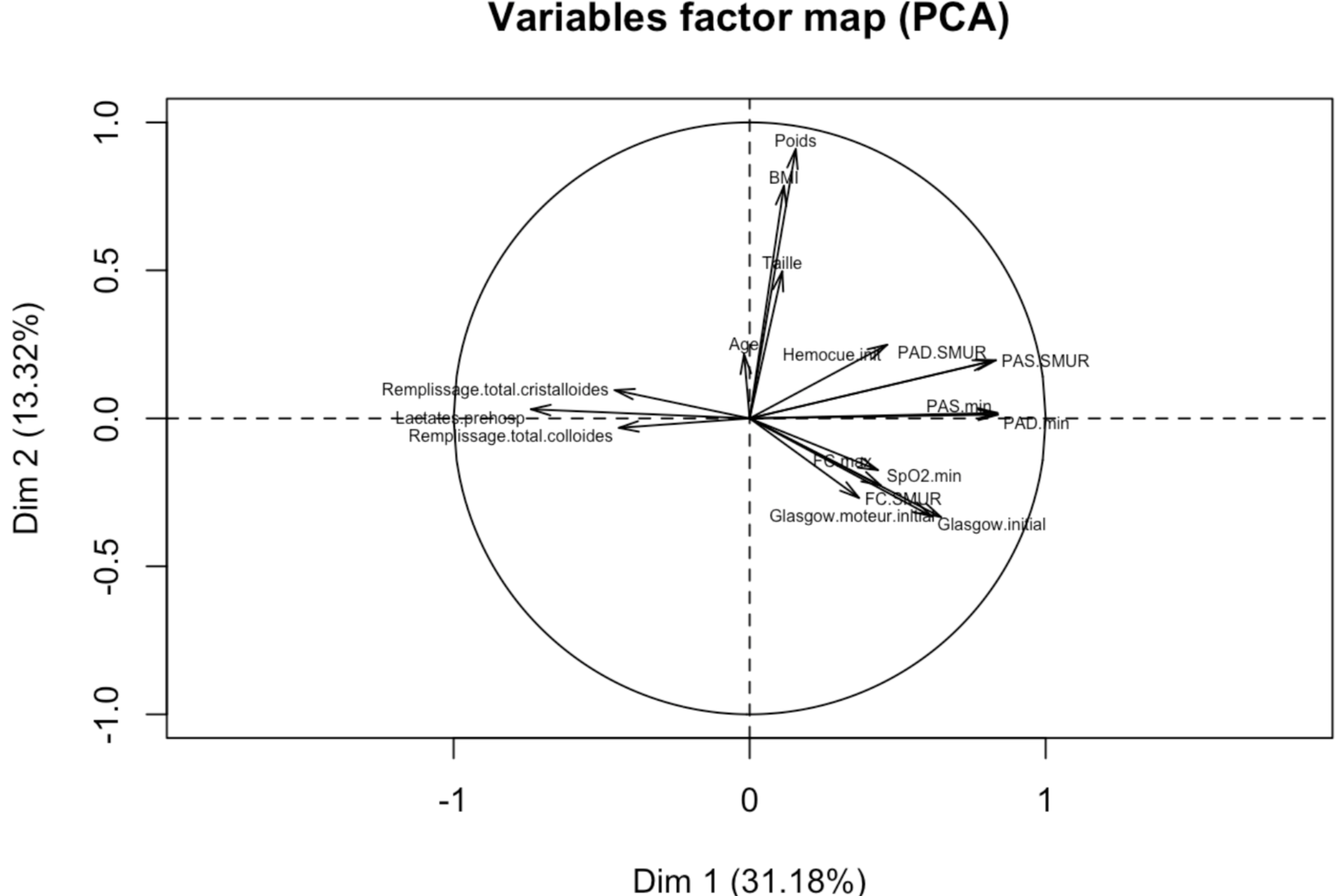}}
\caption{\label{pca}The factor map of the variables from PCA.}
\end{figure} 

There was strong collinearity between variables, as can be seen in the variables PCA factor map (obtained by running an EM-PCA algorithm \citep{missMDA} which performs PCA with missing values) in Figure \ref{pca}, in particular between the minimum systolic (PAS.min) and diastolic blood pressure (PAD.min). 
Based on expert advice, the recoded variables, SD.min and SD.SMUR ($\text{SD.min}=\text{PAS.min} -\text{PAD.min}$; $\text{SD.SMUR}=\text{PAS.SMUR} -\text{PAD.SMUR}$) were used  since they have more clinical significance \citep{redflag}. Thus, we had 14 variables to predict hemorrhagic shock.
\begin{figure}[!htbp]
\centering
    \makebox{
\includegraphics[width=0.95\textwidth]{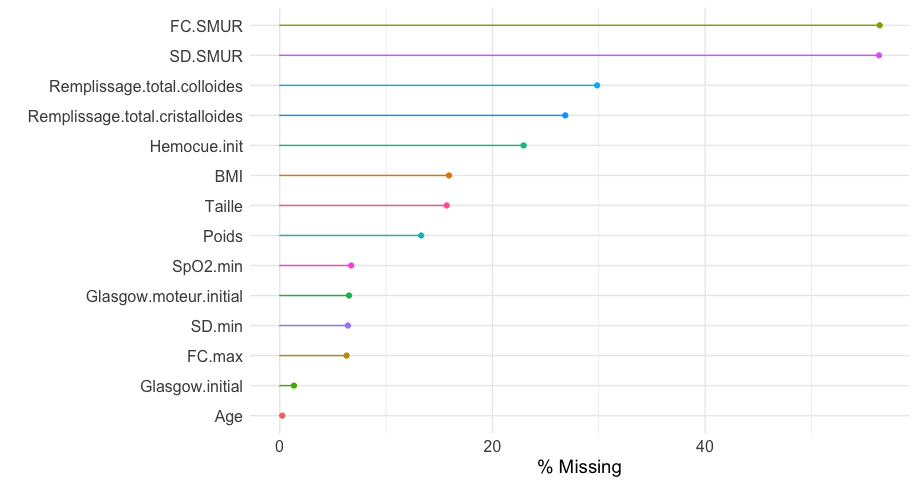}}
\caption{\label{pct_miss} Percentage of missing values in each variable.}
\end{figure} 

Figure \ref{pct_miss} shows the percentage of missingness per variable, varying from  0 to 60\%, which demonstrates the importance of taking appropriate account of missing data.
Even though,  there may be many reasons why missingness occurred, in the end, considering them all to be MAR remains a plausible assumption. 
For instance, FC.SMUR (heart rate) and SD.SMUR  (the difference between blood pressure measured when the ambulance arrives at the accident site) contain many missing values because doctors collected these data during transportation. However, many other medical institutes and scientific publications used measurement on arrival at the accident scene.  
Consequently, doctors decided to record these measures as well but after the TraumaBase was set up.

\begin{figure}[!htbp]
\centering
    \makebox{
\includegraphics[width=0.8\textwidth]{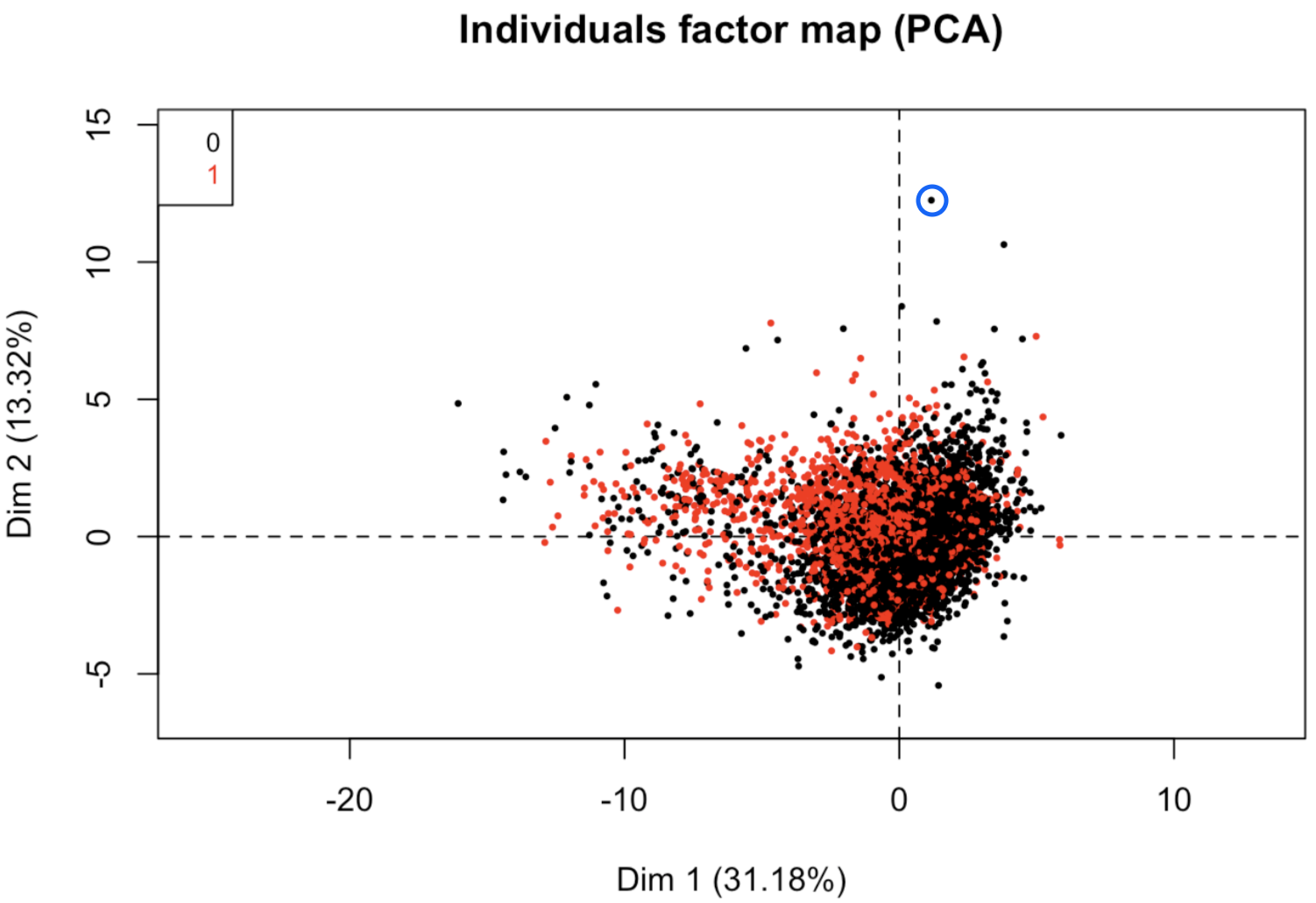}}
\caption{\label{pca_ind} Observation's factor map of PCA. Red points are hemorrhagic shock patients, and black points are patients who did not have hemorrhagic shock. Patient number 3302  (circled in blue) has wrong calculation of BMI.}
\end{figure}

We first applied SAEM for logistic regression with all 14 predictors and for the whole dataset. The estimation obtained by SAEM was of the same order of magnitude as that obtained by multiple imputation. 
Next, we used the model selection procedure described in Section \ref{sectionmodselect}. There were two observations leading to a very small value of the log-likelihood. Upon closer inspection, we found that for patient number $3302$, the  BMI was obtained using an incorrect calculation, and for patient number $1144$, the  weight (200 kg) and height (100 cm) values were likely to be incorrect. Hence, the observed log-likelihood allowed us to discover undetected outliers. On the observations' map of PCA, as shown in Figure \ref{pca_ind}, patient number 3302  (circled in blue) is one of such outliers.
\subsection{Predictive performances}\label{subsection:pred}
We divided the dataset into training and test sets. The training set contained a random selection of 70\% of observations, and the test set contained the remaining 30\%. In the training set, we selected a model with the suggested BIC with missing values, and used forward selection resulting in a model with 8 variables. The estimates of parameters and their standard errors are shown in Table \ref{table:est}.
\begin{table}[!htbp]
\caption{\label{table:est} Estimation of $\beta$ and its standard errors obtained by SAEM, using BIC for model selection.}
\centering
\fbox{%
\begin{tabular}{lr}
  Variables &  Estimate (standard errors)\\ 
  \hline
  $(Intercept)$ & -0.52 (0.59) \\ 
  $Age$ & 0.011 (0.0033) \\
  $Glasgow.moteur$ & -0.16 (0.036)\\         
  $FC.max$ & 0.026 (0.0025)\\
  $Hemocue.init$ & -0.23 (0.031)\\           
 $RT.cristalloides$ & 0.00090 (0.00010) \\
 $RT.colloides$ & 0.0019 (0.00021) \\    
 $SD.min$ & -0.025 (0.0050)\\             
 $SD.SMUR$ & -0.021 (0.0056)\\ 
\end{tabular}}
\end{table}

The TraumaBase medical team indicated to us that the signs of the coefficients were in agreement with their a priori ideas: all the others things being equal \emph{a)} Older people are more likely to have a hemorrhagic shock; \emph{b)} A low Glasgow score implies little or no motor response, which often is the case for hemorrhagic shock patients; \emph{c)} One typical sign of hemorrhagic shock is rapid heart rate; \emph{d)} The more a patient bleeds, the lower their Hemocue is, and the more blood must be transfused. Eventually, it is more likely they will end up in hemorrhagic shock; \emph{e)} Therapy involving two types of volume expanders, cristalloides and colloides, can be conducted to treat hemorrhagic shock. If extremely low difference between blood pressure is observed, its cause may be low stroke volume, as is usually the case in  hemorrhagic shock.

Next, we assessed the  prediction quality on the test set with usual metrics based on the confusion matrix (false positive rate, false negative rate, etc.). 
We need to ensure that the cost of a false negative is much more than that of a false positive, as non-recognition of a potential hemorrhagic shock leads to a higher risk of patient mortality. 
We define the validation error on test set as: 
\begin{equation}\label{eq::cost}
\begin{split}
l(\hat{y},y) = \frac{1}{n} \sum_{i=1}^n w_0 {\mathbbm{1}}_{\{y_i =1, \hat{y}_i=0\}}+ w_1 {\mathbbm{1}}_{\{y_i =0, \hat{y}_i=1\}}
\end{split}
\end{equation}
where $w_0$ and $w_1$ are user defined weight for the cost of false negative and false positive respectively, s.t.,$w_0+w_1=1$.
Therefore, we can choose a threshold for logistic regression by given the value for $w_0$ and $w_1$. For instance, we chose $\frac{w_0}{w_1} = 5$, i.e., the false negative was 5 times costly than the false positive. The cost function was chosen in agreement with the experts. 
\textcolor{black}{
Note that the test set was also incomplete, so we used the strategy described in Subsection \ref{sectionpredna}.}
The confusion matrix of the predictive performance on the test set is shown in Table \ref{table:conf}. The associated ROC curve is shown in Figure \ref{fig:ROC}, and the AUC is 0.8487. 
\begin{table}[!htbp]
\small
\begin{minipage}[b]{0.45\linewidth}
\centering
\begin{tabular}{c >{\bfseries}r @{\hspace{0.7em}}c @{\hspace{0.4em}}c @{\hspace{0.7em}}l}
  \multirow{10}{*}{\parbox{0.5cm}{\bfseries\raggedleft \rotatebox{90}{Observed value} }} & 
    & \multicolumn{2}{c}{\bfseries Predicted outcome} & \\
  & & 1 & 0 \\
  & 1 & \MyBox{True \\Positive}{($98$)} & \MyBox{False \\Negative}{($29$)}  \\[2.4em]
  & 0 & \MyBox{False \\Positive}{($146$)} & \MyBox{True \\Negative}{($808$)} \\ 
  \hspace{0.4em}
\\[3.4em]
 \end{tabular}
\caption{Confusion matrix for prediction on test set.}
 \label{table:conf}
\end{minipage}\hfill
\begin{minipage}[b]{0.45\linewidth}
\centering
\includegraphics[width=1\textwidth]{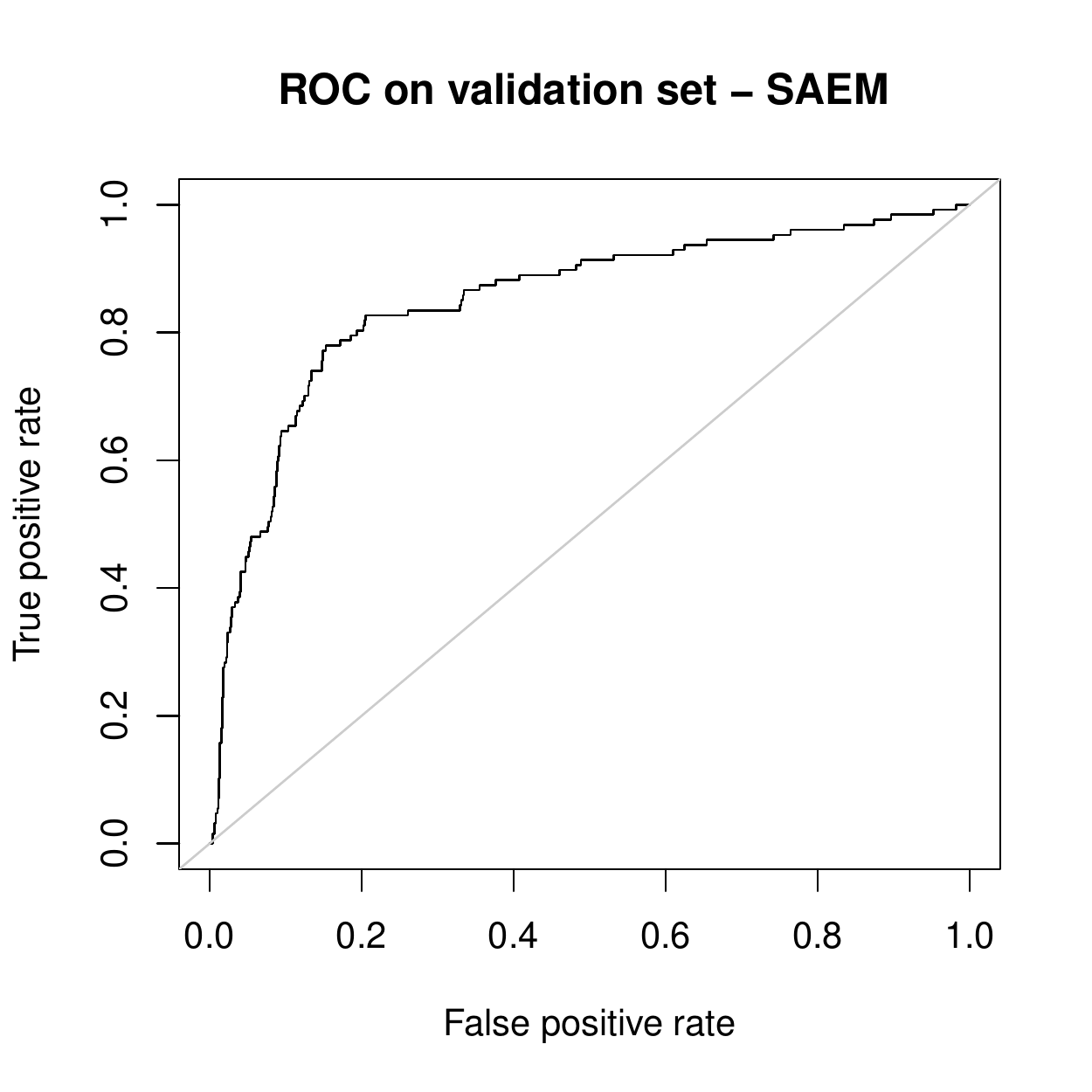}
\captionof{figure}{ROC curve of the test set predictions.}
\label{fig:ROC}
\end{minipage}
\end{table}


\subsection{Comparison with other approaches}
Finally we compared the proposed method to other approaches. \textcolor{black}{Similar to the Subsection \ref{subsection:pred}, we considered  single imputation methods followed by classical logistic regression and variable selection on the imputed training dataset,} such as single imputation by PCA (impPCA) \citep{missMDA}, imputation by Random Forest (missForest) \citep{missforest}, as well as mean imputation (impMean). Meanwhile, we compared logistic regression model with other prediction models, such as Random Forest (predRF) and  SVM (predSVM), both applied on the imputed dataset by Random Forest \citep{missforest}. 
We also considered multiple imputation by chained equation (mice):
we applied logistic regression with a classical forward selection method, with BIC on each imputed data set. However, note that there is no straightforward solution for combining multiple imputation and variable selection; we followed the empirical approach suggested in \citet{model_select_mice}, where they kept the variables selected in each imputed dataset to define the final model.

We also considered three rules used by the doctors to predict the hemorrhagic shock \emph{i)} Doctors' prediction (doctor): the decision was recorded in the TraumaBase. It determines whether the doctor considered the patient to be at risk of hemorrhagic shock.  \emph{ii)} Assessment of Blood Consumption score (ABC):  it is an examination usually performed when the patient arrives at the trauma center. As such, the score is not exactly prehospital but can be computed very early once the patient is hospitalized. \emph{iii)} Trauma Associated Severe Hemorrhage score (TASH): this score was also designed for hemorrhage detection, but at a later stage since it uses some values that are only available after laboratory tests or radiography.
 
 \begin{figure}[!htbp]
\begin{center}
\includegraphics[width=0.9\textwidth]{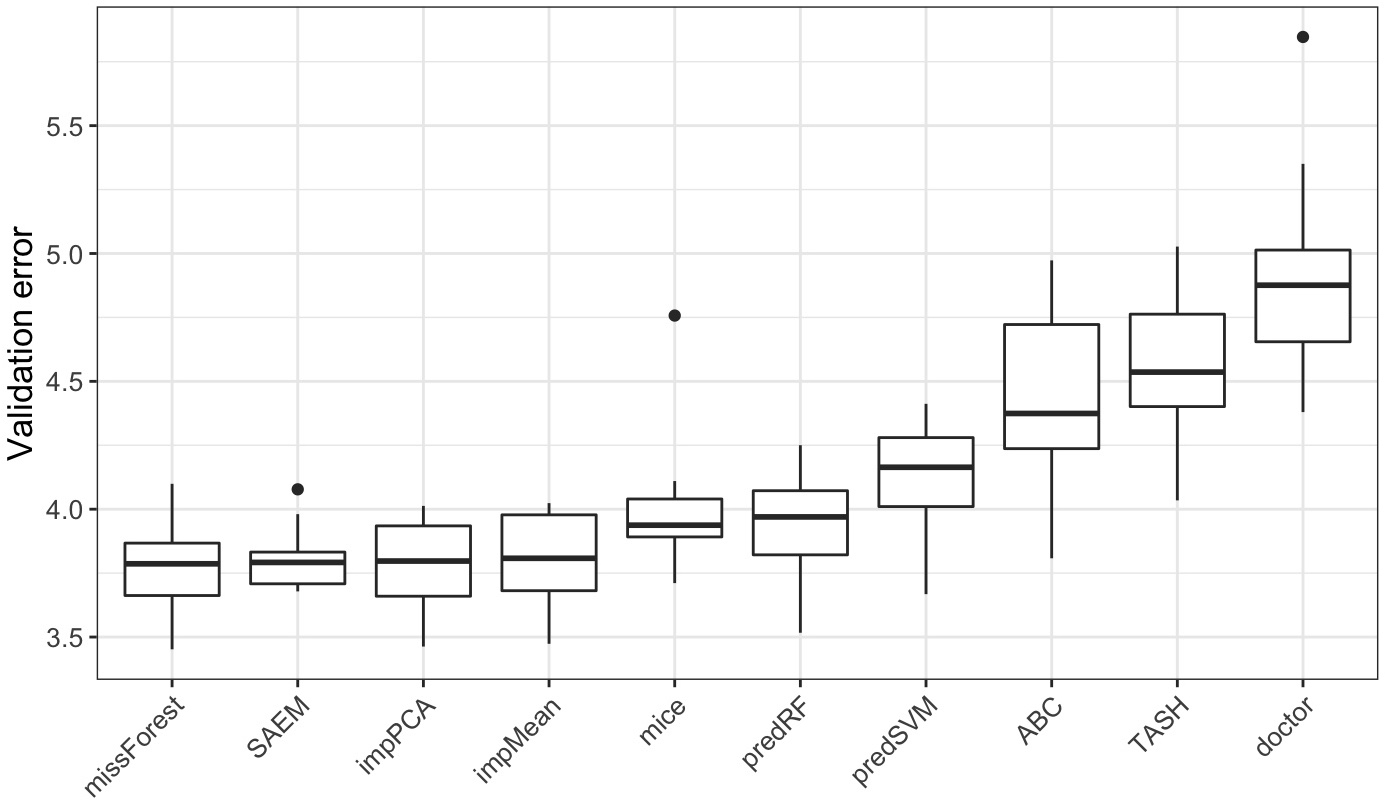}
\caption{Empirical distribution of prediction errors of different methods over 15 replications for the TraumaBase data.}
\label{box::error}
\end{center}
\end{figure}  

Figure \ref{box::error} compares the methods in terms of their  validation error (\ref{eq::cost}). The splitting of data (into training and test sets) was repeated 15 times and we fixed the threshold such that the cost of false negative is 5 times that of false positive, i.e, $\frac{w_0}{w_1}=5$. \textcolor{black}{On average, SAEM had good performance with small variability}, while all the imputation methods performed similarly even the naive mean imputation.  In addition, other prediction methods (Random Forest and SVM) did not result in a smaller error on the test sets than the logistic regression models. Lastly the rules used by the doctors, even the ones using more information than prehospital data,  were not as competitive as SAEM. \ref{ann:tbl} gives the details with classical measures  (AUC, sensitivity, specificity, accuracy and precision) to compare the predictive performance of the methods. The suggested approach resulted in good performance on average, and in particular, had an advantage in terms of the sensitivity, i.e., it rarely misdiagnosed the hemorrhagic shock patients, which is relevant to clinical needs of emergency doctors.

\begin{figure}[!htbp]
\begin{center}
\includegraphics[width=0.9\textwidth]{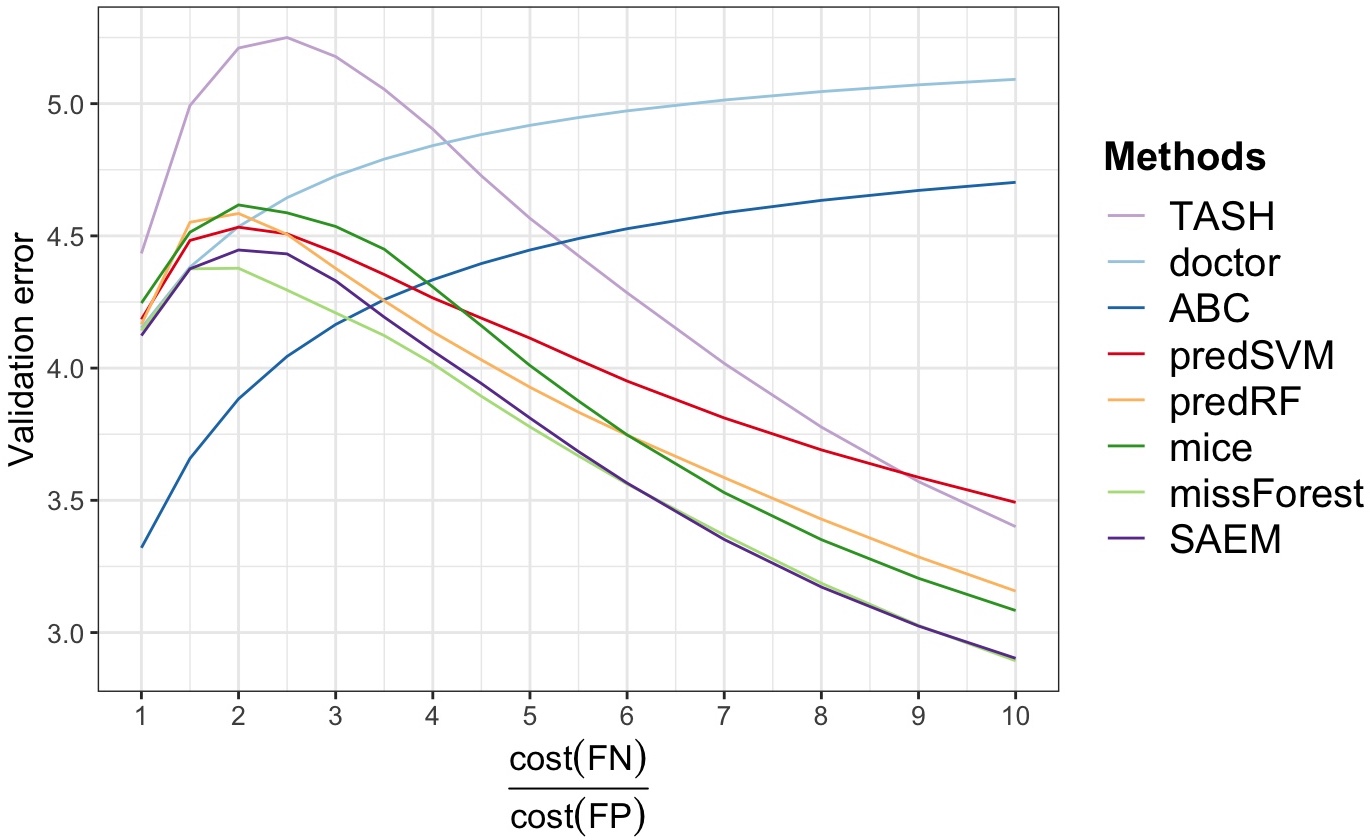}
\caption{Average prediction errors of different methods, as function of the cost importance $\{\frac{w_0}{w_1} \mid \frac{w_0}{w_1}>1 \}$, over 15 replications for the TraumaBase data.}
\label{line::error}
\end{center}
\end{figure} 

More generally, without defining a specific threshold, we observed in Figure \ref{line::error} the average predictive loss over 15 replications as function of the cost importance $\{\frac{w_0}{w_1} \mid \frac{w_0}{w_1}>1 \}$ for all the methods. Obviously, we had the same performance evaluation as before, as SAEM had smaller error on the test sets  with the respect to the choice of $\frac{w_0}{w_1}$, especially when we emphasized more on the cost of false negative. Note that the curves of doctors' rules and ABC increase as a function of the cost importance $\frac{w_0}{w_1} $, which means that, the rules of doctors are more conservative than SAEM, which can be problematic in this application.

Note that even if the proposed methodology is based on the assumption of normally distributed covariates, the performance of the proposed methodology is better than the prediction made by the widely used medical criterion, in terms of prediction error.  Some discussions on the normal assumption are provided in \ref{ann:trauma}.

In summary, the logistic regression methodology with missing values, from estimation to selection, as well as prediction on a test sample with missing data, is theoretically well founded. Based on the TraumaBase application and comparison with other methods, we have demonstrated that the proposed approach has the ability to outperform existing popular methods dealing with missing data.

\section{Discussion}\label{sectiondisc}

In this paper, we have developed a comprehensive joint-modeling framework for logistic regression with missing values.  The experiments indicate that the proposed method is computationally efficient, and can be easily implemented. In addition, compared with multiple imputation  -- especially in the case with correlation between variables -- estimation using SAEM is less biased than other methods and generally leads to interval-estimate coverage that is close to the nominal level. Based on the proposed algorithm, model selection by BIC with missing data can be  performed in a natural way. In view of the results reported in this article, we have been invited by emergency-room doctors in one of the centers that contributes to the TraumaBase dataset to implement the missing-data methodology outlined here in a prospective study to evaluate its performance in real time in a clinical setting. 
Paths for possible future research include further developing the method to handle quantitative and categorical data. 
This paper focused on making inference with missing values but we have suggested a method to predict from a test set with missing values.  More work can be done in the direction of supervised learning with missing values, especially to suggest variance of prediction. Extensions of the methods of \citet{schafer2000inference} could be studied. 
In addition, in the TraumaBase dataset, we can reasonably expect to have both MAR and missing not at random (MNAR) values. MNAR means that missingness is related to the missing values themselves, therefore, the correct treatment would require incorporating models for the missing data mechanisms.
As a final note, the proposed method may be quite useful in the causal inference framework, especially for  propensity score analysis, which estimates the effect of a treatment, policy, or other intervention.  Indeed, inverse probability weighting methods (IPW) are often performed with logistic regression, and the proposed method offers a potential  solution for times where there are missing values in the covariates. The method is implemented in the R package \textit{misaem}.


\newpage
\appendix

\section{Appendix}\label{app}

\subsection{Missing mechanism}\label{ann:ignorable}
Missing completely at random (MCAR) means that there is no relationship between the missingness of the data and any values, observed or missing. 
In other words, MCAR means:
\begin{equation*} \label{mcar}
\dens(M_i|y,\bx_i, \phi)=\dens(M_i|\phi)
\end{equation*}
Missing at Random (MAR), means that the probability to have missing values may depend on the observed data, but not on the missing data.
We must carefully define what this means in our case by decomposing the data $x_i$ into a subset $\xa_i$ of data that ``can be missing'', and a subset $\xb_i$ of data that ``cannot  be missing'', i.e. that are always observed.
Then, the observed data $\xiobs$ necessarily includes the data that can be observed $\xb_i$, while the data that can be missing $\xa_i$ includes the missing data $\ximis$. Thus, MAR assumption implies that, for all individual $i$,
\begin{equation*} \label{mar}
\begin{split}
\dens(M_i | y_i,\bx_i; \phi) &= \dens(M_i | y_i,\xb_i; \phi) \\ &= \dens(M_i | y_i,\xiobs; \phi)
\end{split}
\end{equation*}

MAR assumption implies that, the observed likelihood can be maximize and the  distribution of $M$ can be ignored \citep{little_rubin}. Indeed, 
\begin{equation*} 
\begin{split}
\like(\theta,\phi ; y,\xobs,M)&= \dens(y,\xobs,M ; \theta,\phi) \\
&=\prod_{i=1}^n \dens(y_i,\xiobs,M_i ; \theta,\phi)\\
& = \prod_{i=1}^n \int \dens(y_i,\bx_i,M_i ; \theta,\phi)d\ximis\\
& = \prod_{i=1}^n \int \dens(y_i,\bx_i;\theta) \dens(M_i|\by_i,\bx_i;\phi)d\ximis\\
& = \prod_{i=1}^n \int \dens(y_i,\bx_i;\theta) \dens(M_i|\by_i,\xiobs;\phi)d\ximis\\
& = \prod_{i=1}^n \dens(M_i|\by_i,\xiobs;\phi) \times \prod_{i=1}^n \int \dens(y_i,\bx_i;\theta)d\ximis\\
&= \dens(M|y,\xobs;\phi) \times \dens(y,\xobs; \theta) \\
&= \dens(M|y,\xb;\phi) \times \dens(y,\xobs; \theta) \\
\end{split}
\end{equation*}
Therefore, to estimate $\theta$, we aim at maximizing $\like(\theta; y,\xobs)=\dens(y,\xobs; \theta)$.

\subsection{Metropolis-Hastings sampling}\label{ann:mh}
During the iterations of SAEM, the Metropolis-Hastings sampling is performed as Algorithm \ref{alg1}, with the target distribution $f(\ximis)=\dens(\ximis|\xiobs,\by_i;\theta)$ and the proposal distribution $g(\ximis)=\dens(\ximis|\xiobs;\mu,\Sigma)$. 
\begin{algorithm}[!htbp]
\caption{Metropolis-Hastings sampling.} 
\label{alg1}
\begin{algorithmic}

\REQUIRE An initial samples $\ximis^{(0)}\sim g(\ximis)$;
\FOR{$s = 1,2,\cdots,S$} 
\STATE Generate $\ximis^{(s)}\sim g(\ximis)$; 
\STATE Generate $u \sim \mathcal{U}[0,1]$; 
\STATE Calculate the ratio $w= \frac{f(\ximis^{(s)})/g(\ximis^{(s)})}{f(\ximis^{(s-1)})/g(\ximis^{(s-1)})}$;

\IF{$u<w$} 
\STATE Accept $\ximis^{(s)}$;
\ELSE
\STATE $\ximis^{(s)} \leftarrow \ximis^{(s-1)}$;
\ENDIF 
\ENDFOR 
\ENSURE $(\ximis^{(s)}, 1\leq i \leq n, 1 \leq s \leq S)$.
\end{algorithmic}
\end{algorithm}

\subsection{Calculation of observed information matrix}\label{ann:var}
Procedure \ref{ag2} shows how we calculate the observed information matrix.

\begin{algorithm}[!htbp]
\floatname{algorithm}{Procedure}
\caption{Calculation of observed information matrix.} 
\label{ag2}
\begin{algorithmic}
\REQUIRE After drawing MH samples $(\ximis^{(s)}, 1\leq i \leq n, 1 \leq s \leq S)$ for unobserved data $(\ximis, 1\leq i \leq n)$, we have imputed observations, noted as $(\bz_i^{(s)}, 1\leq i \leq n, 1 \leq s \leq S)$, where $\bz_{ij}^{(s)} = \xiobs, \text{ if } x_{ij} \text{ is observed}; \text{ else } \bz_{ij}^{(s)} =\ximis^{(s)}$. 
\FOR{$n = 1,2,\cdots,n$} 
\FOR{$s = 1,2,\cdots,S$} 
\STATE Calculate the gradient:\\
$\nabla f_{is}=\frac{\partial \llike(\theta;\xiobs,\ximis^{(s)},y_i)}{\partial\bbeta}=
\bz_i^{(s)}\left(y_i-\frac{\exp(\hat{\beta}_0 + \sum_{j=1}^p \hat{\beta}_j\bz_{ij}^{(s)})}{1+\exp(\hat{\beta}_0 +\sum_{j=1}^p \hat{\beta}_j\bz_{ij}^{(s)})}\right)$;
\STATE Calculate the Hessian matrix:\\
$H_{is}=\frac{\partial^2 \llike(\theta;\xiobs,\ximis^{(s)},y_i)}{\partial \bbeta \partial \bbeta^T}= 
-\bz_i^{(s)}{\bz_i^{(s)}}^T \frac{\exp(\hat{\beta}_0 +\sum_{j=1}^p \hat{\beta}_j\bz_{ij}^{(s)})}{\left(1+\exp(\hat{\beta}_0 +\sum_{j=1}^p \hat{\beta}_j\bz_{ij}^{(s)})\right)^2}$;
\STATE $\Delta_i \leftarrow \frac{1}{s}[(s-1)\Delta_i+\nabla f_{is}]$;
\STATE $D_i \leftarrow \frac{1}{s}[(s-1)D_i+H_{is}]$;
\STATE $G_i \leftarrow \frac{1}{s}[(s-1)G_i+\nabla f_{is}\nabla f_{is}^T]$;
\ENDFOR
\STATE $\hat{\mathcal{I}}_S(\hat{\bbeta})\leftarrow\hat{\mathcal{I}}_S(\hat{\bbeta}) -(D_{i}+G_i-\Delta_i \Delta_i^T)$;
\ENDFOR
\ENSURE $\hat{\mathcal{I}}_S(\hat{\bbeta})$.
\end{algorithmic}
\end{algorithm}

\subsection{Logistic regression on simulated complete dataset}
\label{ann:orig}

Figure \ref{fig:rocorig} shows the ROC curve on a simulated complete dataset. The corresponding AUC (for training set) is 0.8976.


\subsection{Simulation results for Missing at Random data}\label{ann:mar}

\begin{figure}[!htbp]
   \begin{minipage}{0.49\textwidth}
     \centering
\includegraphics[width=0.8\textwidth]{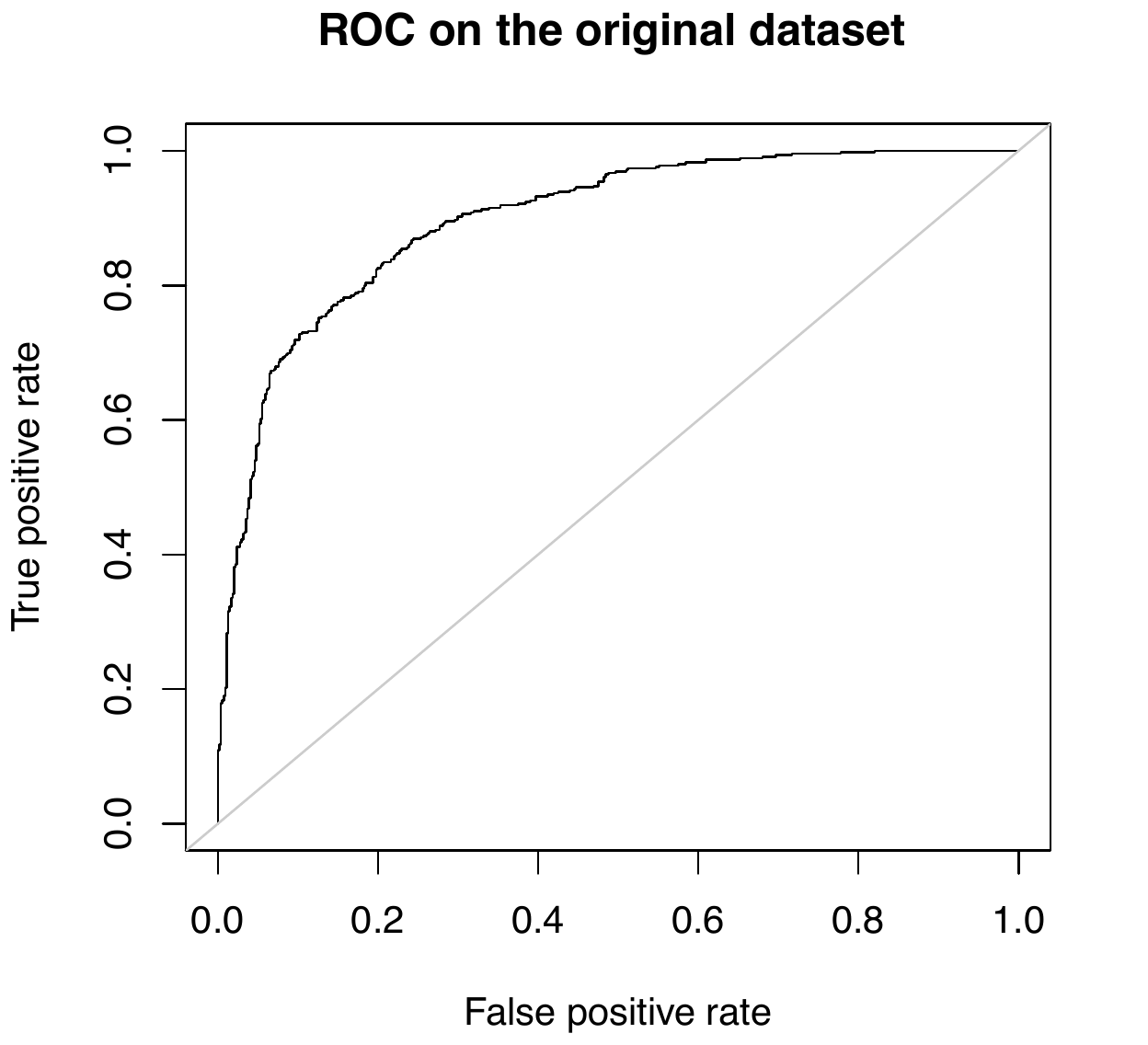}
\caption{ROC curve on a simulated complete dataset.}
\label{fig:rocorig}
   \end{minipage}\hfill
      \begin{minipage}{0.49\textwidth}
     \centering
  \makebox{
\includegraphics[width=0.9\textwidth]{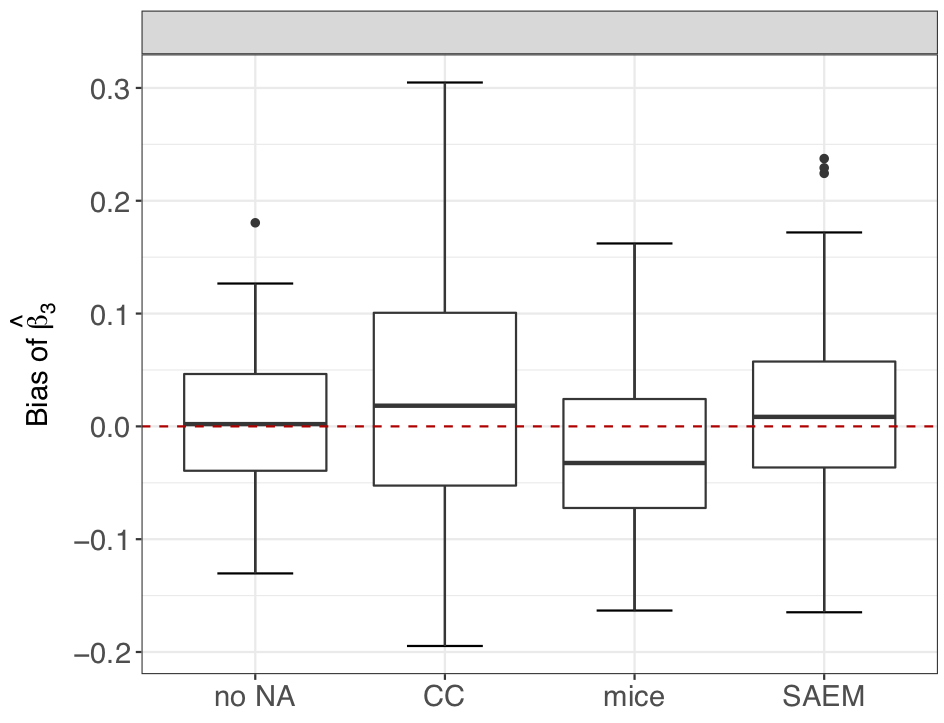}}
\caption{\label{fig:mary}Empirical distribution of the bias of $\hat{\beta}_3$ obtained under MAR mechanism, with $n=1000$ and 10\% of missing values.}
   \end{minipage}
\end{figure}

We consider a Missing at Random mechanism to generate data. Figure \ref{fig:mary} shows that the biases were very similar to the ones obtained under a MCAR mechanism and the parameters were estimated without bias. 
\vspace{-0.2cm}
\vspace{-0.2cm}
\subsection{Simulation results for model misspecification: the coverage} \label{ann:miscoverage}
Table \ref{mis:coverage} shows the coverage for all parameters and inside the parentheses is the average length of corresponding confidence interval. 
\vspace{-0.2cm}
\begin{table}[!htbp]
\caption{\label{mis:coverage} Coverage (\%) for $n=1000$, MCAR and misspecified models, calculated over 1000 simulations. Bold indicates under coverage. Inside the parentheses is the average length of corresponding confidence interval over 1000 simulations (multiplied by 100).}
\centering
\fbox{
\begin{tabular}{lcccr}%
  parameter & no NA & CC & mice & SAEM \\ 
  \hline
  Student distribution:   & ($v=5$)& & &\\
  $\beta_0$ & 94.7 (68.02) & 94.3 (84.14) & 94.6 (67.69) & 93.8 (68.25)\\
    $\beta_1$ & 95.2 (54.78) & 94.2 (72.15) & 91.7 (61.96) & 93.5 (63.05) \\
      $\beta_2$ & 94.9 (27.66) & 94.6 (36.39) & 91.4 (31.21) & 93.7 (31.84)\\
        $\beta_3$ & 94.9 (26.76) & 94.3 (35.24) & \textbf{81.5} (30.46) & 94.7 (29.98)\\
          $\beta_4$ & 95.2 (11.52) & 95.4 (15.16) & 95.8 (12.94) & 95.5 (12.88)\\
            $\beta_5$ & 93.7 (17.63) & 94.9 (23.22) & \textbf{83.4} (20.40) & 93.3 (19.93)\\
              \hline
  Gaussian mixture:   & & & &\\
  $\beta_0$ & 94.8 (57.54) & 95.2 (75.42)  & 95.4 (61.95) & 95.0 (61.33)\\
    $\beta_1$ & 94.7 (58.00) & 96.2 (76.05) & 95.4 (66.66) & 95.3 (66.13) \\
      $\beta_2$ & 94.3 (28.49) & 95.3 (37.35) & 95.3 (32.65) & 94.0 (32.50)\\
        $\beta_3$ & 94.7 (26.16) & 94.9 (34.38) & 94.9 (28.91) & 94.5 (29.10)\\
          $\beta_4$ & 94.4 (12.68) & 94.4 (16.60) & 94.4 (14.24) & 94.7 (14.09)\\
            $\beta_5$ & 95.3 (17.70) & 94.7 (23.25) & 94.7 (19.86) & 95.3 (19.92)
\end{tabular}
}
\end{table}
\subsection{Definition of the variables of the TraumaBase data set}\label{ann:trauma}
In this Subsection, we give the detailed explanations for the selected quantitative variables:
\begin{itemize}
\item $Age$: Age. 
\item $Poids$: Weight.
\item  $Taille$:  Height.
\item  $BMI$:  Body Mass index, $BMI = \frac{Weight \text{ in } kg}{(Height \text{ in } m)^2}$
\item  $Glasgow$: Glasgow Coma Scale .
\item  $Glasgow.moteur$: Glasgow Coma Scale motor component.
\item  $PAS.min$: The minimum systolic blood pressure. 
\item  $PAD.min$: The minimum diastolic blood pressure. 
\item  $FC.max$:  The maximum number of heart rate (or pulse) per unit time (usually a minute).
\item  $PAS.SMUR$: Systolic blood pressure at arrival of ambulance.
\item  $PAD.SMUR$: Diastolic blood pressure at arrival of ambulance.
\item  $FC.SMUR$: Heart rate at arrival of ambulance.
\item  $Hemocue.init$: Capillary Hemoglobin concentration.
\item  $SpO2.min$: Oxygen saturation.
\item  $Remplissage.total.colloides$ (or $RT.colloides$): Fluid expansion colloids.
\item  $Remplissage.total.cristalloides$ (or $RT.cristalloides$): Fluid expansion cristalloids.
\item $SD.min$ ($=PAS.min -PAD.min$): Pulse pressure for the minimum value of diastolic and systolic blood pressure.
\item  $SD.SMUR$ ($=PAS.SMUR -PAD.SMUR$): Pulse pressure at arrival of ambulance.
\end{itemize}

\begin{figure}[!htbp]
\centering
\subfloat[Histograms of covariates]{\includegraphics[width=0.5\textwidth]{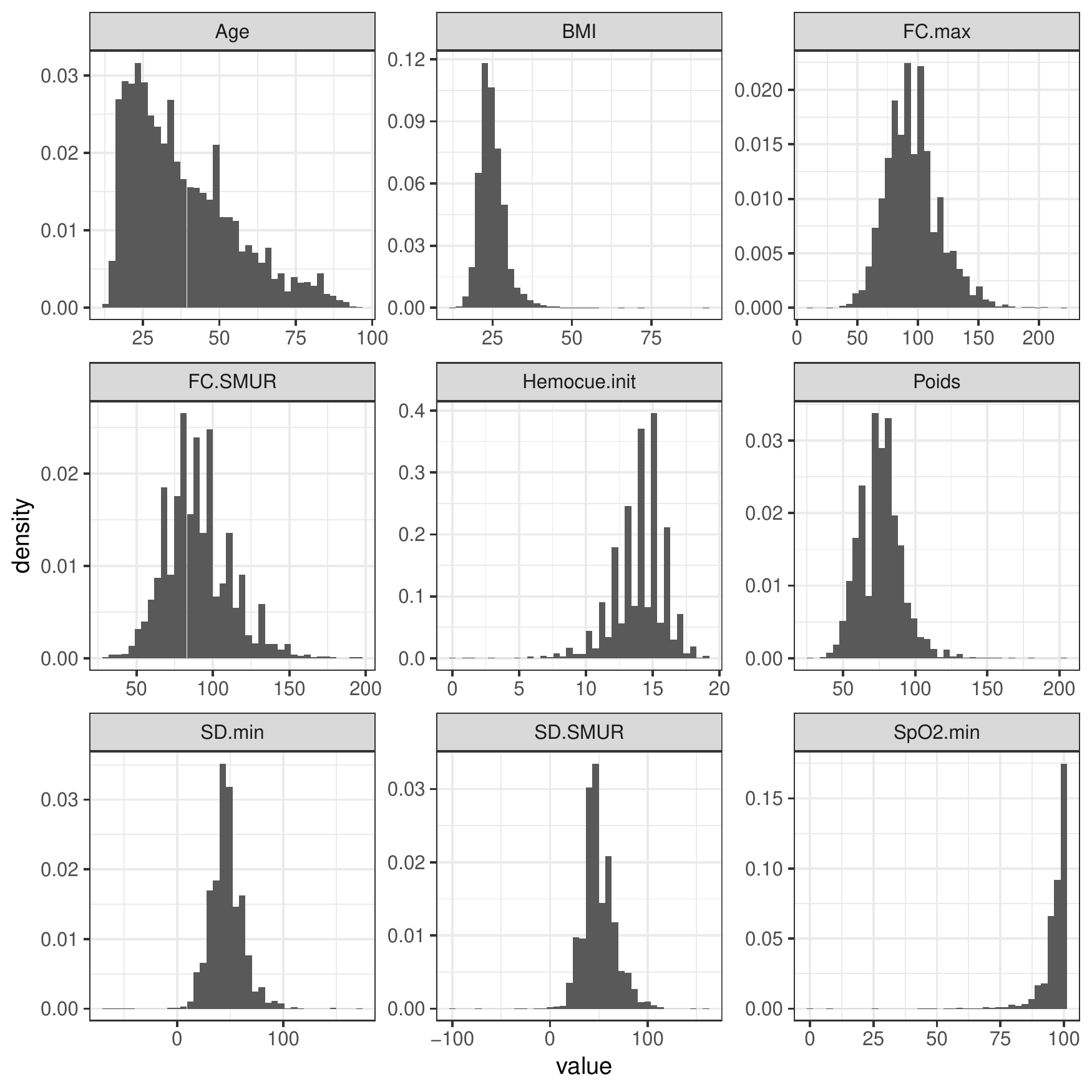}}
\hfill
\subfloat[Empirical cumulative distributions]{\includegraphics[width=0.5\textwidth]{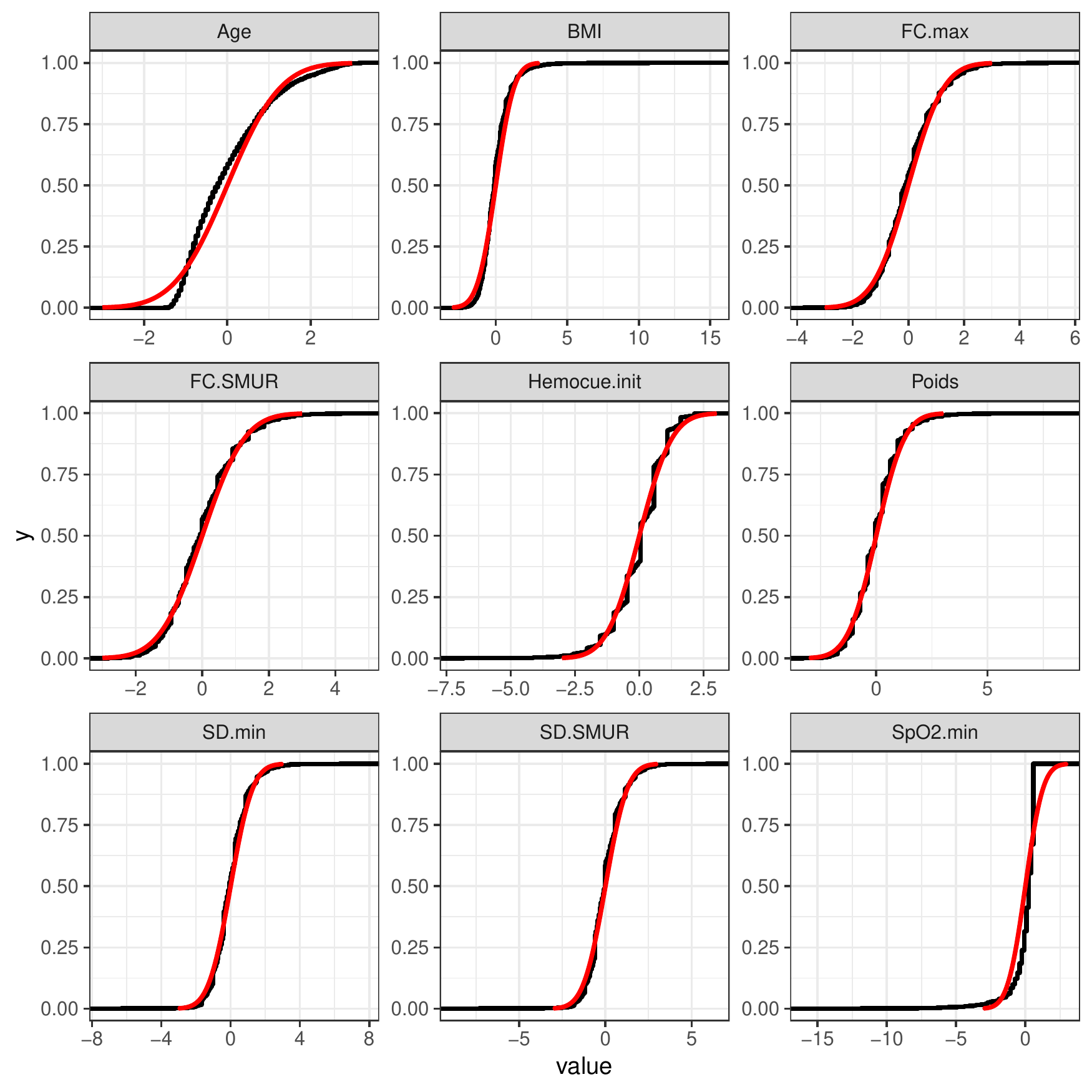}}
\hfill
\caption{\label{fig:plotvar}Empirical distribution of variables from TraumaBase. (a) Histograms of covariates (b) Black curve illustrates the empirical cumulative distributions while the red curve represents the normal distribution.}
\end{figure}

Figure \ref{fig:plotvar} shows the histogram and the empirical \textit{c.d.f.} of several covariates from the TraumaBase data.
Several of these distributions are not symmetrical. In practice, it is possible to consider that some suitable transformations of the covariates can be approximated by normal distributions. For example, transformations of the form  $\log(c+x)$ and $\log(c-x)$, can be very appropriate for, respectively, right-skewed and left-skewed distributions. 
We applied the proposed methodology to the real dataset after transformation.  However, the prediction result from cross-validation didn't show advantage of the transformed version. Indeed when the log transformation is used as a prepossessing step, it only operates on the observed part, which is appropriate under MCAR calues. Consequently, taking into account the simulation study, the interpretability, the choices of transformations, and the prediction results,  we have decided to keep the variables without any transformation. 

\subsection{Details of predictive performance for TraumaBase data}\label{ann:tbl}
Details of predictive performance  for TraumaBase data are given by Table \ref{table:com}.
\begin{table}[!htbp]
\caption{\label{table:com} Comparison of the mean of the predictive performances (values are multiplied by 100) of different methods dealing with missing data. 
AUC is the area under ROC; the accuracy is the number of true positive plus true negative divided by the total number of observations; the sensitivity is defined as the true positive rate; specificity as the true negative rate; the precision is the number of true positive over all positive predictions. The best results are in bold.}
\centering
\fbox{%
\begin{tabular}{lccccccr}
  Metrics & SAEM  & missForest & impMean & impPCA& mice & predRF & predSVM\\ 
    \hline
   AUC & \textbf{88.5} & \textbf{88.8} & \textbf{88.9} & \textbf{89.0} & 87.7 & 88.0 & 80.4\\
  Accuracy &  86.9 & 87.0 &  87.3 & 86.7& 85.3 &87.2 & \textbf{88.3}\\
  Precision & 41.1 & 41.6 &  42.2 & 41.0 &37.9 & 41.6 & \textbf{44.0}\\
Sensitivity & \textbf{74.6} & 74.3 & 73.2 & \textbf{75.0} & \textbf{75.2} & 71.5 & 66.0\\         
  Specificity & 88.2  & 88.4 & 88.8 & 87.9 & 86.4 & 88.9 & \textbf{90.6}\\
\end{tabular}}
\end{table}

\section*{Supplementary material}
      \begin{description}
\item[R-package:] R-package ``misaem'' containing the implementation of algorithm SAEM to fit the logistic regression model with missing data, now available in CRAN \citep{misaem}.
\item[Codes:] Code to reproduce the experiments are provided in GitHub \citep{github}.
\item[Additional supplementary materials:] Some supplementary simulation results are presented \citep{supp}.
\end{description}

\section*{References}
\bibliographystyle{elsarticle-num-names}
\bibliography{bibliography}

\end{document}